\documentclass[useAMS,usenatbib]{mn2e}
\usepackage{longtable}
\usepackage{graphicx}
\usepackage{lscape}
\usepackage{amsmath}
\usepackage{rotating}
\newcommand{\mnras}{MNRAS}
\newcommand{\aj}{AJ}
\newcommand{\apj}{ApJ}
\newcommand{\apjs}{ApJ}
\newcommand{\apjl}{ApJ}
\newcommand{\aap}{A\&A}

\newcommand{\nar}{NewAR}
\newcommand{\nat}{Nature}
\newcommand{\memsai}{Memorie della Societa Astronomica Italiana}
\newcommand{\araa}{Annual Review of Astronomy and Astrophysics}

\newcommand{\Teff}{\mbox{$T_{\mathrm{eff}}$}}

\title[DA white  dwarfs from  the LSS-GAC] {DA  white dwarfs  from the
  LSS-GAC survey  DR1: the  preliminary luminosity and  mass functions
  and formation rate}

\author[A.  Rebassa-Mansergas et al.]{A.
  Rebassa-Mansergas$^{1}$\thanks{Email;
    arebassa@pku.edu.cn; LAMOST Fellow}, X.-W.
  Liu$^{1,2}$\thanks{Email; x.liu@pku.edu.cn}, R.  Cojocaru$^{3,4}$,
  H.-B.  Yuan$^{1}$, S.  Torres$^{3,4}$,\newauthor
  E. Garc\'ia-Berro$^{3,4}$, M.-X.  Xiang$^{2}$, Y.  Huang$^{2}$,
  D. Koester$^{5}$, Y. Hou$^{6}$, G. Li$^{7}$, Y. Zhang$^{6}$\\
$^{1}$  Kavli   Institute  for  Astronomy  and   Astrophysics,  Peking
  University, Beijing 100871, P.\,R.\,China\\
$^{2}$Department of Astronomy, Peking University, Beijing 100871,
  P.\,R.\,China\\
$^{3}$ Departament de F\'\i sica Aplicada, Universitat Polit\`ecnica
  de Catalunya, c/Esteve Terrades 5, 08860 Castelldefels,
  Spain\\ 
$^{4}$ Institute for Space Studies of Catalonia, c/Gran Capit\`a 2--4,
  Edif. Nexus 104, 08034 Barcelona, Spain\\
$^{5}$ Institut f\"ur Theoretische Physik und Astrophysik, University
  of Kiel, 24098 Kiel, Germany\\ 
$^{6}$ Nanjing Institute of Astronomical Optics \& Technology,
  National Astronomical Observatories, Chinese Academy of Sciences,\\
  Nanjing 210042, P.\,R.\,China\\
$^{7}$ Key Laboratory of Optical Astronomy, National Astronomical
  Observatories, Chinese Academy of Sciences, Beijing 100012,
  P.\,R.\,China\\
}

\begin{document}
\date{Accepted 2015. Received 2015; in original form 2015}
\pagerange{\pageref{firstpage}--\pageref{lastpage}} \pubyear{2015}
\maketitle

\begin{abstract}
Modern large-scale  surveys have  allowed the identification  of large
numbers  of  white dwarfs.   However,  these  surveys are  subject  to
complicated  target   selection  algorithms,  which  make   it  almost
impossible to quantify to what  extent the observational biases affect
the  observed populations.   The LAMOST  (Large Sky  Area Multi-Object
Fiber Spectroscopic  Telescope) Spectroscopic  Survey of  the Galactic
anti-center  (LSS-GAC)  follows a  well-defined  set  of criteria  for
selecting  targets for  observations.   This  advantage over  previous
surveys  has  been  fully  exploited  here to  identify  a  small  yet
well-characterised  magnitude-limited  sample  of  hydrogen-rich  (DA)
white dwarfs.  We derive preliminary LSS-GAC DA white dwarf luminosity
and mass functions. The space density and average formation rate of DA
white dwarfs  we derive are $0.83\pm0.16  \times10^{-3}$ pc$^{-3}$ and
$5.42  \pm  0.08  \times10^{-13}$ pc$^{-3}$  yr$^{-1}$,  respectively.
Additionally, using an existing  Monte Carlo population synthesis code
we simulate the  population of single DA white dwarfs  in the Galactic
anti-center, under various assumptions.  The synthetic populations are
passed through the LSS-GAC selection criteria, taking into account all
possible observational biases.  This allows us to perform a meaningful
comparison of the observed and  simulated distributions.  We find that
the LSS-GAC  set of  criteria is highly  efficient in  selecting white
dwarfs  for  spectroscopic observations  (80-85  per  cent) and  that,
overall,  our  simulations  reproduce  well  the  observed  luminosity
function.   However, they  fail at  reproducing an  excess of  massive
white  dwarfs present  in  the observed  mass  function.  A  plausible
explanation  for this  is that  a  sizable fraction  of massive  white
dwarfs  in the  Galaxy  are  the product  of  white dwarf-white  dwarf
mergers.
\end{abstract}


\section{Introduction}
\label{s-intro}

White dwarfs  (WD) are the typical  endpoint of the evolution  of most
main sequence stars.  Because nuclear  reactions do not occur in their
deep interiors, the evolution of WDs can be considered as a relatively
simple    and   well    understood   gravothermal    cooling   process
\citep{althausetal10-1}. Actually, the  evolutionary cooling times are
now very accurate \citep[e.g.][]{renedoetal10-1}, providing a reliable
way of measuring  the WD cooling age from the  temperature and surface
gravity measured observationally.  WDs are  hence very useful tools in
astronomy.   For  example,  the  WD luminosity  function  (LF)  is  an
important  statistical instrument  which  has been  used  not only  to
derive an accurate age of the  Galactic disk in the solar neighborhood
\citep{wingetetal87-1,  garcia-berroetal88-1,   fontaineetal01-1}  but
also to constrain the  local star formation rate \citep{noh+scalo90-1,
  diaz-pintoetal94-1,  rowell13-1}.  Moreover,  the  WD mass  function
(MF) has been  successfully employed over the years as  a tool to test
the theory of stellar evolution,  offering information on stellar mass
loss. This function  has also been used to study  how the evolution of
close    binaries     is    able     to    produce     low-mass    WDs
\citep{liebertetal05-1}, thus helping to asses the contribution of the
distinct evolutionary scenarios in producing the current population of
Galactic binaries in which one of the components is a WD. Finally, the
WD age function  (AF) is also a valuable tool  for constraining the WD
formation history \citep{huetal07-1}.

With  the advent  of modern,  large-scale  surveys such  as the  Sloan
Digital Sky  Survey \citep[SDSS;][]{yorketal00-1} or  the Super-Cosmos
Sky   Survey  \citep{hamblyetal01-1},   the   size   of  the   current
observational WD samples has increased dramatically.  This has allowed
producing  more  accurate  LFs,  MFs  and  AFs  \citep{harrisetal06-1,
  kepleretal07-1,  huetal07-1, degennaroetal08-1,  rowell+hambly11-1}.
However, the major drawback of  most studies is the complicated target
selection algorithms, which incorporate  observational biases that are
almost impossible  to quantify  in numerical  simulations that  aim at
reproducing the ensemble properties of the observed samples.

Significant  observational efforts  have  also allowed  to unveil  the
population  of WDs  within 20\,pc  of the  Sun \citep{holbergetal02-1,
  holbergetal08-1, giammicheleetal12-1}.  This  sample is considerably
less numerous than those mentioned above,  however it is claimed to be
reasonable  complete  and can  therefore  be  considered as  a  volume
(rather than magnitude) limited sample, which suffers effectively from
no selection biases.

The LAMOST (Large Sky Area Multi-Object Fiber Spectroscopic Telescope)
Spectroscopic     Survey      of     the      Galactic     anti-center
\citep[LSS-GAC;][]{liuetal14-1,  yuanetal15-1} follows  a well-defined
selection criteria aiming at providing  spectra for stellar sources of
all colours in  the Galactic anti-center (including WDs)  so that they
can  be studied  in a  statistically meaningful  way. LSS-GAC  started
operations in 2011  and will provide a significantly  larger sample of
WDs than  those within  20\,pc of  the Sun.  In  this paper  we derive
preliminary observed LF, MF and AF  of WDs identified within the first
data  release  of LSS-GAC,  and  use  a state-of-the-art  Monte  Carlo
population synthesis code adapted to the characteristics of the survey
to simulate the  WD population in the Galactic  anti-center.  We apply
the  LSS-GAC selection  criteria to  the simulated  samples, carefully
evaluate all possible observational  biases, and derive synthetic LFs,
MFs  and  AFs.   This  exercise  allows us  to  perform  a  meaningful
comparison between  the outcome  of simulations and  the observational
data.

\section{The LAMOST spectroscopic survey of the Galactic anti-center}
\label{s-gac}

LAMOST  is a  quasi-meridian reflecting  Schmidt telescope  located at
Xinglong   Observing  Station   in   the  Hebei   province  of   China
\citep{cuietal12-1, luoetal12-1}.  The effective aperture of LAMOST is
about 4  meters.  LAMOST  is exclusively  dedicated to  obtain optical
spectroscopy of celestial objects. Each ``spectral plate'' refers to a
focal  surface with  4,000 auto-positioned  optical fibers  to observe
spectroscopic   plus  calibration   targets  simultaneously,   equally
distributed among  16 fiber-fed  spectrographs.  Each  spectrograph is
equipped  with  two  CCD  cameras   of  blue  and  red  channels  that
simultaneously  provide blue  and red  spectra of  the 4,000  selected
targets, respectively.

The  LSS-GAC  is a  major  component  of  the LAMOST  Galactic  survey
\citep{dengetal12-1,  zhaoetal12-1}.  By  selecting targets  uniformly
and  randomly in  ($r$, $g  - r$)  and ($r$,  $r -  i$) Hess  diagrams
\citep{yuanetal15-1},  the  LSS-GAC  main survey  aims  at  collecting
$\lambda\lambda$3,700 -- 9,000 low resolution ($R \sim 1,800$) spectra
for a statistically complete sample  of $\sim 3$\,million stars of all
colours down to  a limiting magnitude of  $r=$17.8\,mag (18.5\,mag for
limited fields),  distributed in a  contiguous sky area of  over 3,400
deg$^2$ centered  on the Galactic anti-center  ($|b| \leq 30^{\circ}$,
$150 \leq  l \leq  210^{\circ}$).  The  simple yet  non-trivial target
selection of the survey makes possible to study the underlying stellar
populations for any given type of target, such as WDs.  In addition to
the main survey, the LSS-GAC also includes a survey of the M\,31/M\,33
area,  targeting hundreds  of  thousands of  objects  in the  vicinity
fields of  M\,31 and M\,33, and  a survey of Very  Bright (VB) plates,
targeting over  a million of  randomly selected very bright  stars ($r
\le 14$\,mag) in the northern hemisphere during bright/grey nights.

Targets for  the LSS-GAC  main survey  are selected  for spectroscopic
follow-up from  the Xuyi Schmidt  Telescope Photometric Survey  of the
Galactic   anti-center   (XSTPS-GAC)   catalogue   \citep{liuetal14-1,
  zhangetal13-2, zhangetal14-1}.   The XSTPS-GAC survey  was initiated
in October 2009  and completed in March 2011. Photometry  in SDSS $g$,
$r$  and $i$  bands was  acquired with  the Xuyi  1.04/1.20\,m Schmidt
Telescope  located  at  the  Xuyi   station  of  the  Purple  Mountain
Observatory, P.\,R.\,China.   The XSTPS-GAC surveyed an  area of about
5,400\,deg$^2$, from ${\rm  RA} \sim 3$ to 9$^{\rm h}$  and ${\rm Dec}
\sim -10$ to +60$^{\rm o}$ to cover the LSS-GAC main survey footprint,
plus an extension of $\sim 900$\,deg$^2$ to the M\,31, M\,33 region to
cover the  LSS-GAC M\,31/M\,33 survey area.   The resulting catalogues
contain about  100 million stars  down to  a limiting magnitude  of $r
\sim 19.0$\,mag ($10\sigma$), with an astrometric calibration accuracy
of   0.1\,arcsec  \citep{zhangetal14-1}   and  a   global  photometric
calibration  accuracy of  $\sim$2 per  cent \citep[][Yuan  et al.,  in
  prep.]{liuetal14-1}.

For the LSS-GAC  main survey, bright (B), medium-bright  (M) and faint
(F) plates  are designated to target  sources of brightness 14.0$  < r
\la$ 16.3\,mag,  16.3 $  \la r \la  $17.8\,mag and 17.8  $ \la  r \leq
18.5$\,mag, respectively.  During the Pilot  (Oct.  2011 -- Jun. 2012)
and the first year Regular (Oct. 2012 -- Jun. 2013) Surveys of LAMOST,
a total  of 1,042,586  [750,867] spectra  of a  signal to  noise ratio
(S/N)  $\ge$ 10  at  7450\AA\, [S/N  $\ge$ 10  at  4650\AA] have  been
collected, including  439,560 [225,522] spectra from  the main survey.
Most of the stars are from the B plates.

The raw  data were reduced with  the LAMOST 2D pipeline  (Version 2.6)
\citep[e.g.][]{luoetal04-1}, following the standard procedures of bias
subtraction,    cosmic-ray    removal,   1D    spectral    extraction,
flat-fielding,  wavelength  calibration,  and  sky  subtraction.   The
LAMOST spectra  are recorded in  two arms, 3,700 --  5,900\,\AA~in the
blue  and 5,700  --  9,000\,\AA~in  the red.   The  blue- and  red-arm
spectra  are processed  independently in  the 2D  pipeline and  joined
together  after flux  calibration.  Flux  calibration for  the LSS-GAC
observations is  performed using  an iterative algorithm  developed at
Peking University  by \citet{xiangetal15-1}, achieving an  accuracy of
about 10 per cent for the whole wavelength ranges of blue- and red-arm
spectra.  A  detailed description  of the target  selection algorithm,
survey design, observations, data  reduction, and value-added catalogs
of    the   first    LSS-GAC    data   release    is   presented    in
\citet{yuanetal15-1}.  In  the current  work we only  consider spectra
obtained by the main LSS-GAC survey. A list including the names of the
plates  employed for  the main  survey is  provided in  Table\,A1 (see
Appendix).

\section{The LSS-GAC WD sample}
\label{s-sample}

\begin{figure}
\begin{center}
\includegraphics[angle=-90,width=\columnwidth] {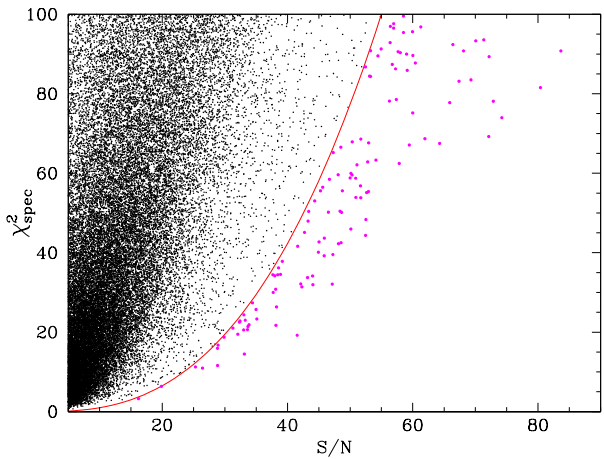}
\caption{\label{f-chi}  Reduced  $\chi^{2}_\mathrm{spec}-\mathrm{S/N}$
  distribution   obtained  by   fitting  one   of  our   WD  templates
  [\Teff=14,000K,  $\log(g)$=7  dex,  see Fig.\,\ref{f-temp}]  to  the
  entire LSS-GAC spectroscopic data base.  Objects falling in the area
  defined by $\chi^2_\mathrm{spec}<0.002\times\mathrm{S/N}^{2.7}$ (red
  curve) are considered WD candidates (magenta dots).}
\end{center}
\end{figure}

In this section we describe our methodology for identifying WDs within
the  LSS-GAC  spectroscopic  data  base.  We  do  this  following  two
independent but  complementary routines.  The first  identifies WDs by
$\chi^2$-template fitting all LSS-GAC  spectra, the second selects WDs
by applying  a well-defined  colour cut  to the  XSTPS-GAC photometric
catalogue.   We also  estimate the  spectroscopic completeness  of the
LSS-GAC WD sample.

\subsection{The $\chi^2$-template fitting method}
\label{s-chi2}

We   use   the   $\chi^2$-template   fitting   method   described   by
\citet{rebassa-mansergasetal10-1}  to  identify   LSS-GAC  WDs.   This
routine was originally  developed to search for  SDSS WD-main sequence
binaries                    \citep[WDMS;][]{rebassa-mansergasetal10-1,
  rebassa-mansergasetal12-1, rebassa-mansergasetal13-2}.   However its
mathematical  prescription can  be easily  implemented to  the LSS-GAC
data  for identifying  single  WDs.  As  the first  step,  a given  WD
template is used to fit the entire LSS-GAC spectroscopic data base and
the resultant reduced $\chi^2_{\rm  spec}$ values are recorded.  Those
$\chi^2_{\rm  spec}$, together  with  the overall  S/N  ratios of  the
LSS-GAC spectra  are represented in  a 2D map  and an equation  of the
form,

\begin{equation}
\chi^2_\mathrm{max}=a\times\mathrm{SN}^{b},
\end{equation}

\begin{figure}
\begin{center}
\includegraphics[angle=-90,width=\columnwidth] {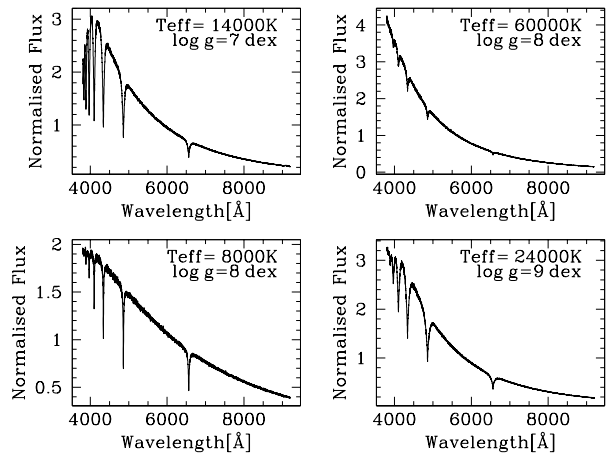}
\caption{\label{f-temp} Four  WD template examples used  in this work.
  Effective temperature and surface gravity  are indicated for each of
  them.}
\end{center}
\end{figure}

\noindent  is   defined  so   that  all   systems  below   this  curve
($\chi^2_\mathrm{spec}<\chi^2_\mathrm{max}$)  are   considered  as  WD
candidates, where  $a$ and  $b$ are free  parameters defined  for each
template.  An  example is illustrated in  Fig.\,\ref{f-chi}.  The form
of  Eq.\,(1)  is defined  to  account  for systematic  errors  between
template and observed spectra, which  become more important the larger
the S/N is.  The particular shape  of the curve is carefully evaluated
for   each  template   as   a  compromise   between   the  number   of
excluded/selected spectra.   In other  words, we  aim at  selecting as
many WD candidate spectra as  possible of similar spectral features as
the  templates  used (note  that,  strictly  speaking, the  only  free
parameter  we   are  considering  in   the  fitting  process   is  the
normalization factor).  However, the number of selected targets cannot
be too large, as otherwise the automatization becomes useless.

The spectra of all WD  selected candidates are then visually inspected
and systems  that are not WDs  are simply excluded.   This exercise is
repeated for all WD templates considered.  In this case, our templates
are  the  result  of  adding   artificial  Gaussian  noise  to  45  DA
(hydrogen-rich)  carefully  selected  WD  model  spectra  provided  by
\citet{koester10-1}  that   cover  a  broad  range   of  WD  effective
temperatures (6,000K-60,000\,K) and surface  gravities (7 $< \log\,g <
$\,9\,dex).  The  model spectra were  binned to the  LSS-GAC resolving
power.    Four  of   our  considered   WD  templates   are   shown  in
Fig.\,\ref{f-temp}.  Obviously,  using only DA  WD templates restricts
our  search   to  hydrogen-rich  WDs.   However,  as   we  show  below
(Section\,\ref{s-compl}), the number  of non-DA WDs currently observed
by the LSS-GAC is rather small.   Moreover, it has to be stressed that
the  presence of  a small  fraction  of non-DA  WDs does  not seem  to
influence significantly the shape of the LF \citep{cojocaruetal14-1}.

As already  mentioned, LSS-GAC spectra  are the result of  combining a
blue and a red optical spectrum obtained from two separate arms. Given
that single WDs  are blue objects, in principle, the  blue arm spectra
alone  should  allow  us  to  efficiently  identify  such  stars.   To
investigate this, we separately applied the above described routine to
all blue-arm  LSS-GAC spectra as  well as  to all combined  (blue- and
red-arm) spectra and  compared the results.  We found  that the number
of identified WDs differed significantly.  Specifically, we identified
$\sim30$  per cent  more objects  using the  combined, blue  plus red,
spectra.  Moreover, all WDs identified using only the blue-arm spectra
were included in the list of  targets found using the combined LSS-GAC
spectra.   The  overall  shape  of the  continuum  spectrum  is  hence
required for efficiently  selecting single WDs.  We  found this effect
to be most important when selecting  WDs against hot (A) main sequence
stars in  which the strengths of  the Balmer lines are  similar.  This
was also the case when selecting WDs amongst low S/N ($\la5$) spectra.
For  these two  reasons,  we  decided to  only  consider the  combined
LSS-GAC spectra  of S/N  ratio above  5 in  this work.   This excluded
15,442   LSS-GAC   spectra  for   which   the   default  2D   pipeline
(Section\,\ref{s-gac}) failed  to deliver usable red-  and/or blue-arm
spectra.   It also  excluded 622,112  spectra of  S/N ratios  below 5.
This  requirement  is further  supported  by  the fact  that  reliable
stellar parameters  (necessary for obtaining  reliable LF, MF  and AF)
cannot    be    derived    for    low    S/N    ratio    WD    spectra
\citep{rebassa-mansergasetal10-1}.  Thus, the  total number of LSS-GAC
spectra  considered in  this work  was 306,600,  of which  20,029 were
selected as WD candidates by  our $\chi^2$-template fitting method.  A
visual inspection revealed that 94 of  those are genuine WDs, 81 of DA
type  (see  Table\,\ref{tab:numbers}).   We also  identified  26  WDMS
binaries.  With  the exception of J080357.21+334138.6,  these binaries
are  already  included   in  the  LAMOST  WDMS   binary  catalogue  by
\citet{renetal13-1, renetal14-1}.

\begin{figure}
\begin{center}
\includegraphics[angle=-90,width=\columnwidth] {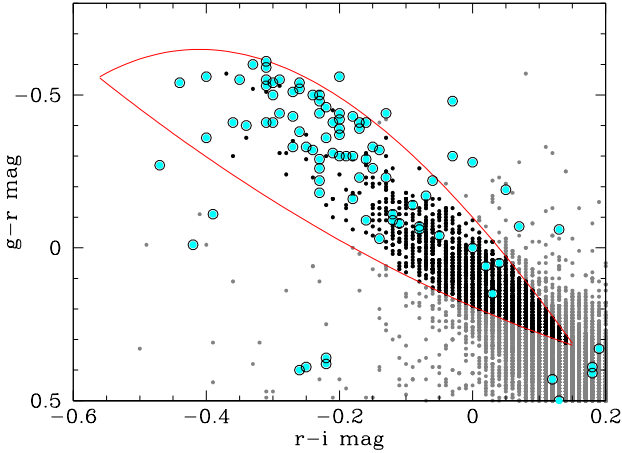}
\caption{\label{f-colcut}  ($g-r$) versus  ($r-i$) colour-colour  plot
  illustrating  the  location of  LSS-GAC  DA  white dwarf  candidates
  (black solid  dots within the  colour cut  defined by the  red solid
  lines). The 92 confirmed LSS-GAC DAs  are shown as cyan dots, others
  as grey  dots The $g$ magnitudes  of 22 LSS-GAC DA  white dwarfs are
  found  to be  unreliable, which  explains  why these  WDs are  found
  outside the expected colour region (two of them beyond the displayed
  area).}
\end{center}
\end{figure}

\subsection{The colour cut method}

In the  previous section we  searched for DA WDs  by $\chi^2$-template
fitting  all  LSS-GAC  spectra.   Here  we  complement  this  with  an
independent strategy  based on XSTPS-GAC  $g$, $r$ and  $i$ photometry
(Section\,\ref{s-gac}).   This exercise  relies simply  on applying  a
well-defined  cut  to  the  XSTPS-GAC $g-r$  and  $r-i$  colours  (see
Table\,1 of \citealt{girvenetal11-1}  and Fig.\,\ref{f-colcut} of this
paper).   \citet{girvenetal11-1} provide  two additional  colour cuts,
using the $u$  and $z$ magnitudes respectively, however  these are not
used here because  the XSTPS-GAC does not provide  magnitudes in those
filters.  16,824 sources  fall within the area defined by  our cut and
visual inspection of  spectra of those targets confirms 78  as WDs, of
which 70 are DAs ---  see Table\,\ref{tab:numbers}.  The vast majority
of the  remaining targets are  single A,  F main sequence  stars whose
colours overlap with those of  cool WDs.  The contamination by quasars
is negligible as  these sources are generaly too faint  to be observed
by the LSS-GAC survey.  63 (57) of these 78 (70) WDs (DAs) are already
discovered  following the  above  described $\chi^2$-template  fitting
method.  Thus, the colour method adds  15 new objects (13 of which are
DAs) to  our sample.  The total  number of LSS-GAC WDs  thus raises to
105,    and    among    them    we   classify    90    as    DA    WDs
(Table\,\ref{tab:numbers}).   The complete  LSS-GAC  DA  WD sample  is
provided in Table\,A2 (see Appendix).

Two  important  conclusions can  be  drawn  from the  above  exercise.
First,  the  $\chi^2$-template  fitting   method  failed  to  identify
$\sim15$  per  cent  of  the  whole  LSS-GAC  DA  WD  sample.   Visual
inspection of  the spectra  of those systems  revealed that  they were
either of low S/N ratios ($\sim5-6$  in the blue-arm spectra), or were
subject to artifacts of a  bad flux calibration/merging of the two-arm
(blue plus  red) spectra,  or a combination  of both.   Although those
effects clearly affect the identification  of DA WDs when applying the
$\chi^2$-template  fitting method,  they do  not influence  the colour
selection of DA WDs.  Secondly, the DA WD colour cut we applied missed
$\sim22$  per cent  of the  whole DA  WD sample  identified. A  closer
inspection  revealed  the  $g$  magnitudes  of  these  targets  to  be
unreliable.  This  caused those  objects to fall  far from  the colour
locus expected for DA WDs (see Fig.\,\ref{f-colcut}).

\begin{table}
\caption{\label{tab:numbers} Number of WDs  (including DAs, DBs, etc.)
  and DAs  alone identified by  the $\chi^2$-S/N template  fitting and
  colour methods, in common by  the two methods, independently by each
  method, and missed by both methods.  The total number of LSS-GAC WDs
  and DAs are provided in the last row.}
\begin{center}
\begin{tabular}{lcc}
\hline
\hline
 &  $N_{\rm WD}$ & $N_{\rm DA}$ \\
\hline
$\chi^2$-S/N method & 94  & 81 \\
Colour method & 78  & 70 \\
Common by both & 63  & 57 \\
$\chi^2$-S/N method alone & 27  & 20 \\
Colour method alone & 15  & 13 \\
Missed by both &  2  &  2 \\
Total & 107 &  92 \\
\hline
\end{tabular}
\end{center}
\end{table}

\begin{figure*}
\begin{center}
\includegraphics[width=0.49\textwidth]{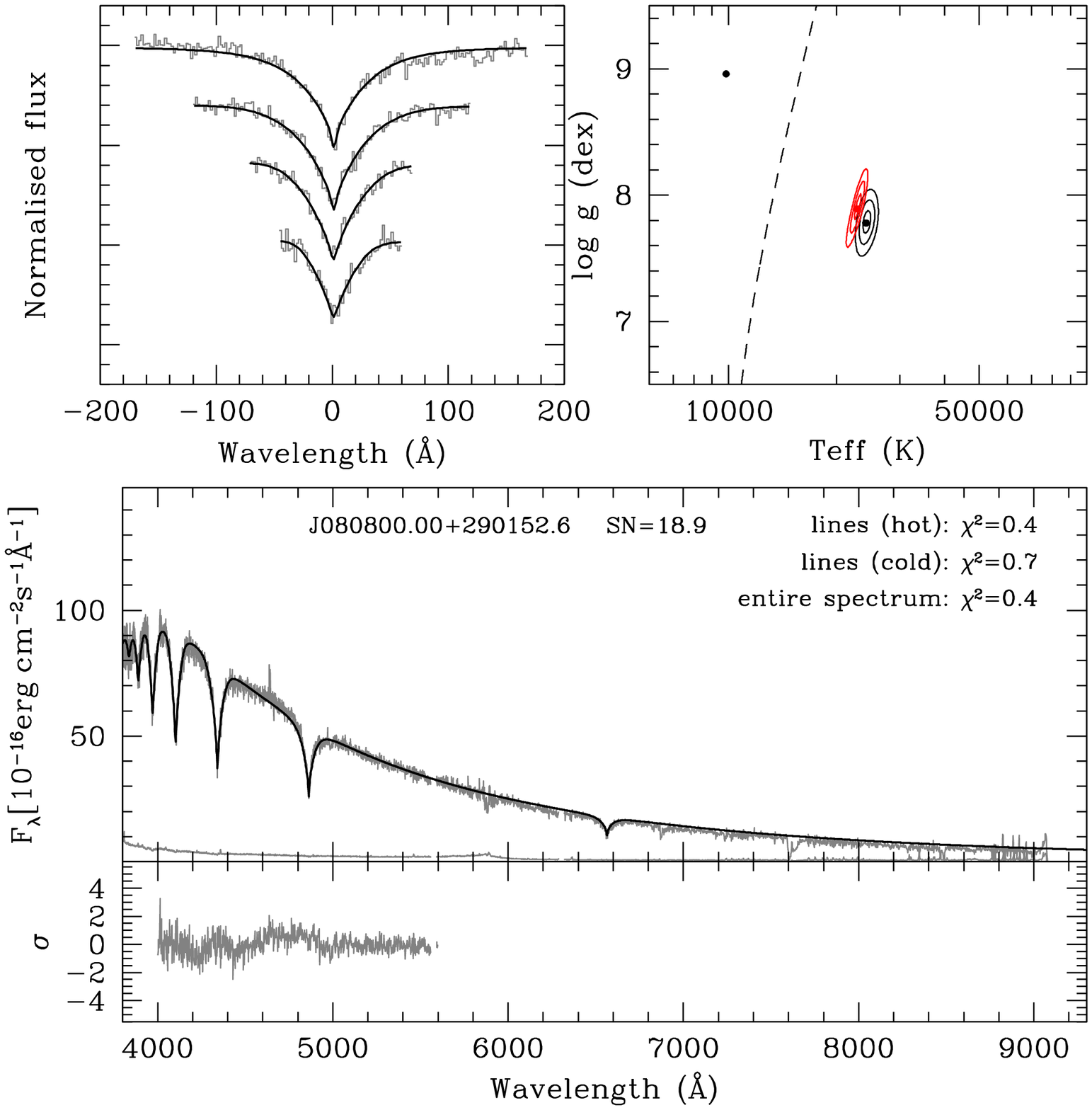}
\hfill
\includegraphics[width=0.49\textwidth]{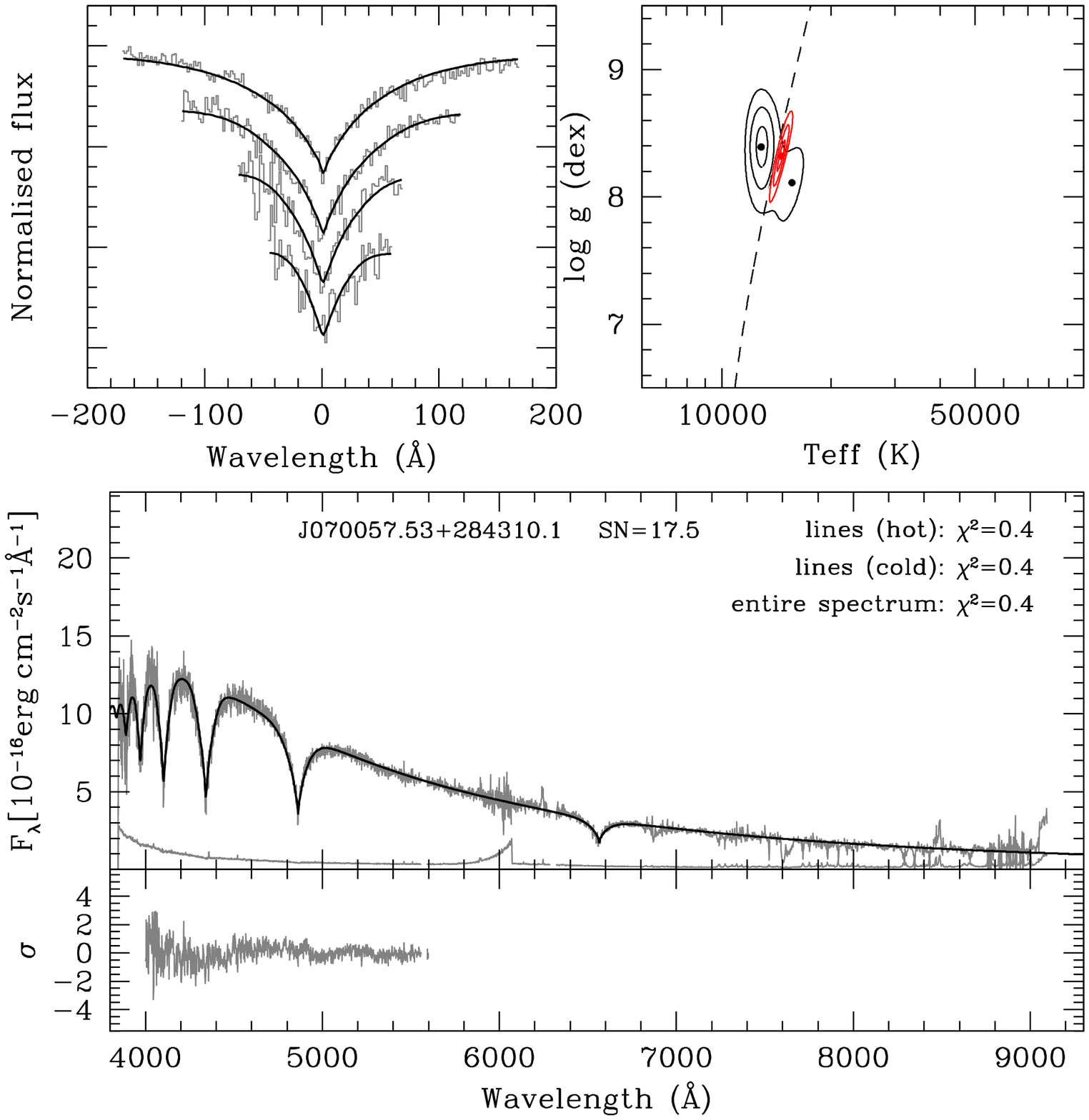}
\caption{\label{f-fits}  Spectral model  fits  to two  DA  WDs in  our
  sample.  {\em Top  row, 1st and 3rd panels}:  best-fit (black lines)
  to  the  normalised H$\beta$  to  H$\epsilon$  (grey lines,  top  to
  bottom) line profiles.  {\em Top row,  2nd and 4th panels}: 1, 3 and
  6$\sigma$ $\chi^2$ contour  plots in the $\Teff-\log  g$ plane.  The
  black contours refer to the best  line profile fit, the red contours
  to  the fit  of the  entire  (continuum plus  lines) spectrum.   The
  dashed line indicates the  occurrence of maximum H$\beta$ equivalent
  width.  The  best ``hot''  and ``cold''  line profile  solutions are
  indicated  by  black dots,  the  best  fit  to the  entire  spectrum
  (continuum plus lines) is indicated by  a red dot.  {\em Bottom row,
    upper panels}:  the WD spectra and  associated uncertaintied (grey
  lines) along with  the best-fit model spectrum (black  line) for the
  4,000--5,500\,\AA\ wavelength  range. The $\chi^2$ that  result from
  the  Balmer line  fitting (cold  and hot  solutions) and  the entire
  spectrum  fitting  are  also  indicated.   {\em  Bottom  row,  lower
    panels}.      The    residuals     of    the     fit    in     the
  4,000--5,500\,\AA\  wavelength  range.   The $\Teff$  and  $\log  g$
  values listed in Table\,A2 are  determined from the best Balmer line
  profile fit.  The fit to the  entire spectrum is only used to select
  between the  ``hot'' and ``cold''  solutions of Balmer  line profile
  fit (see also Fig.\,\ref{f-solutions}).}
\end{center}
\end{figure*}

\subsection{Spectroscopic completeness of the  LSS-GAC DA WD sample}
\label{s-compl}

 The spectroscopic completeness  is defined as the fraction  of DA WDs
 that we have identified compared to  the total number of WDs observed
 by  the LSS-GAC.   It can  be assumed  that the  DA WD  spectroscopic
 sample  is 100  per cent  complete  within the  colour selection  box
 defined by the $g-r$ and $r-i$ colours, as we have visually inspected
 every single  spectrum within this  region.  There  are 70 DA  WDs in
 this  colour box,  of which  57 were  found by  the $\chi^2$-template
 fitting   method   (Table\,\ref{tab:numbers}).    The   spectroscopic
 completeness  of  the  $\chi^2$  method  within  the  colour  box  is
 therefore  80  per   cent.  The  number  of  DA  WDs   found  by  the
 $\chi^2$-template  fitting  method  outside  the  colour  box  is  20
 (Table\,\ref{tab:numbers}).   If  we  assume that  the  spectroscopic
 completeness of the $\chi^2$-template  fitting method is not strongly
 colour dependent,  then the total number  of DA WDs we  expect to lie
 outside the colour  box is 20/0.8 =  25.  The total number  of DA WDs
 that could  have been  detected is  therefore (70+25)  = 95,  and the
 overall  spectroscopic completeness  is  (70+20)/95  $\simeq$ 95  per
 cent.  The two  independent methods we have used  to identify LSS-GAC
 DA WDs,  namely the $\chi^2$-template  fitting method and  the colour
 cut method,  seem to complement  each other  very well and  manage to
 identify the vast majority of DA WDs observed by the LSS-GAC survey.

\begin{figure*}
\begin{center}
\includegraphics[angle=-90,width=0.49\textwidth]{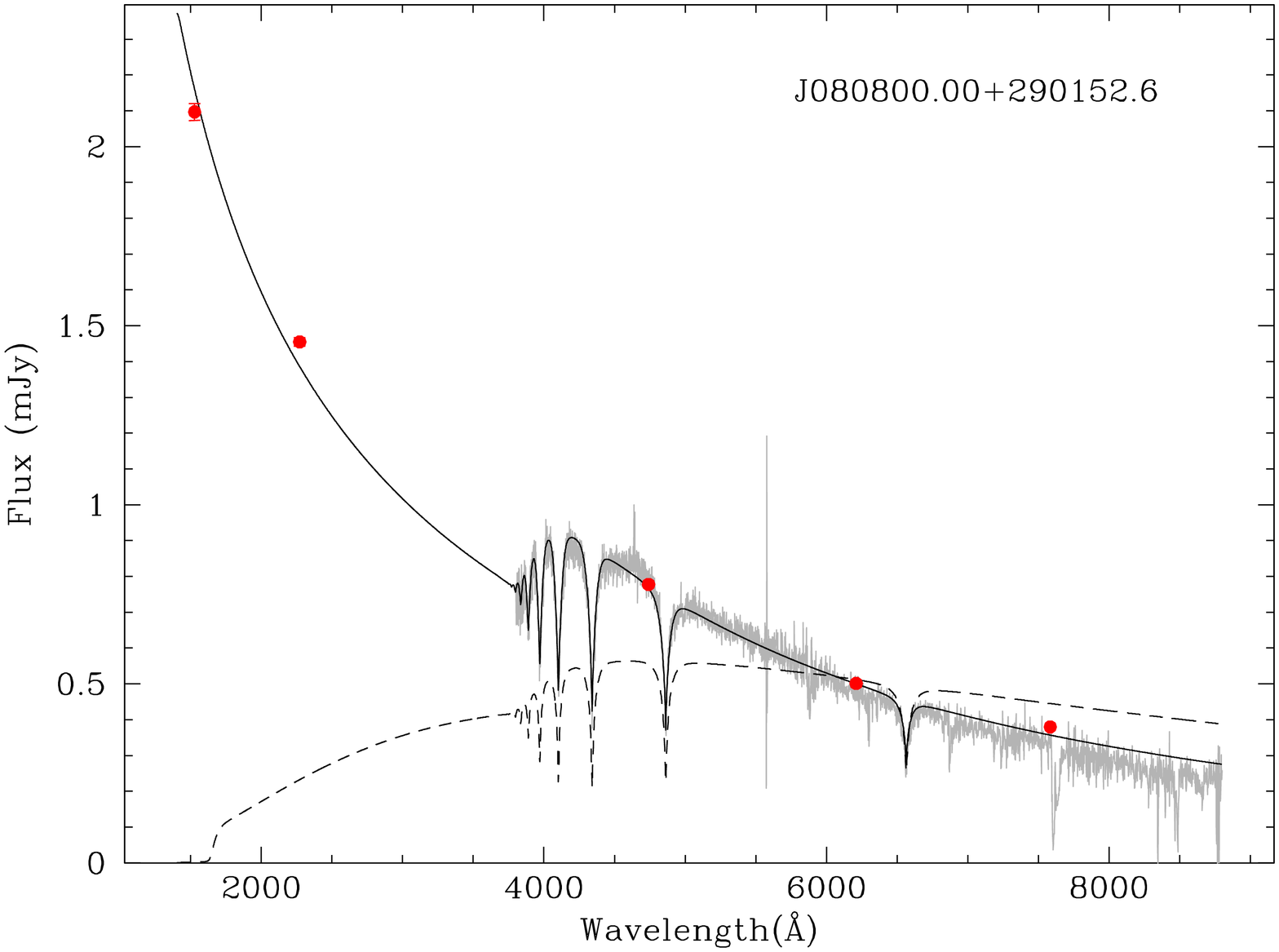}
\hfill
\includegraphics[angle=-90,width=0.49\textwidth]{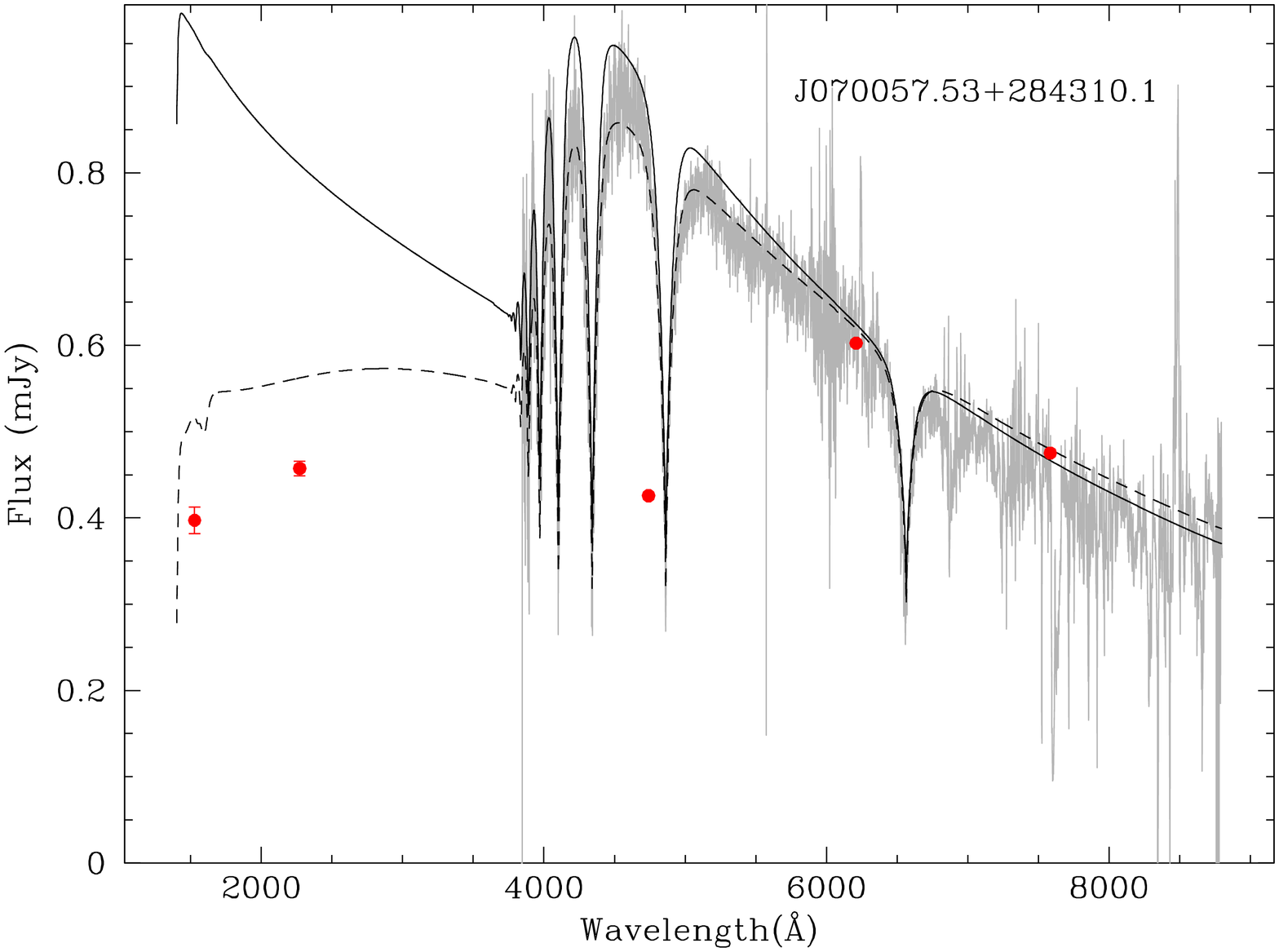}
\caption{\label{f-solutions}. Observed  LSS-GAC spectra of the  two DA
  WDs  shown  in Fig.\,\ref{f-fits}  (gray),  together  with the  best
  Balmer  line model  fits (black  solid line  for the  hot solutions,
  black  dashed line  for the  cold solutions).   The GALEX  near- and
  far-ultraviolet  fluxes and  the XSTPS-GAC  optical fluxes  (derived
  from  the $g,r,i$  magnitudes) are  shown as  red solid  dots.  Left
  panel: the  predicted ultraviolet and optical  fluxes clearly select
  the  hot solution  of  the  Balmer fitting,  in  agreement with  the
  solution   preferred  by   fitting  the   entire  LSS-GAC   spectrum
  (Fig.\,\ref{f-fits}, top row, second  panel). Right panel: the GALEX
  ultraviolet fluxes are  in good agreement with the  cold Balmer line
  fit solution and thus clearly rule  out the hot solution favoured by
  the fit to the entire  LSS-GAC spectrum (Fig.\,\ref{f-fits}, top row
  fourth panel). Note  that in this particular case  the $g$ XSTPS-GAC
  magnitude is found to be unreliable (Fig.\,\ref{f-colcut}).}
\end{center}
\end{figure*}

An  addional way  to  quantify the  spectroscopic  completeness is  by
cross-correlating our list with the  LAMOST DA WD catalogues published
by \citet{zhaoetal13-1} and \citet{zhangyyetal13-1}.  Of the 16 DA WDs
of \citet{zhaoetal13-1},  and 27 of \citet{zhangyyetal13-1}  that fall
in the area observed by LSS-GAC and have spectra of S/N ratios $\geq5$
in both the  red- and the blue-arm  spectra, 15 and 26  are within our
sample, respectively\footnote{In passing we note that the total number
  of objects  from the catalogue by  \citet{zhangyyetal13-1} that fall
  within the Galactic anti-center area is 71, however we exclude 44 as
  they  are  not  WDs.}.   Therefore,   we  have  missed  two  DA  WDs
(Table\,\ref{tab:numbers}).   This exercise  suggests a  spectroscopic
completeness of $\sim94-96$ per cent for our sample, in good agreement
with the above estimated value.

In Table\,A2  we include  the two  DA WDs that  we have  missed.  This
increases our LSS-GAC WD catalogue to 107 WDs. Of these, 92 are of the
DA type  (Table\,\ref{tab:numbers}).  Given that  DA WDs are  the most
common    ones   --    $\sim$85    per   cent,    see   for    example
\cite{kleinmanetal13-1} -- and our final sample contains 92 DA WDs and
seems to be highly complete, it can  be said that the number of non-DA
WDs currently observed by LSS-GAC is very small.

\section{Stellar parameters and distances}
\label{s-param}

We  determine the  stellar  parameters following  the fitting  routine
described by \citet{rebassa-mansergasetal07-1}.   This relies on using
the entire  model grid  of DA  WDs of  \citet{koester10-1} to  fit the
normalised H$\beta$ to  H$\epsilon$ line profiles of  each WD spectrum
for  determining  the  effective  temperature  ($\Teff$)  and  surface
gravity    ($\log   g$).     Two   examples    are   illustrated    in
Fig.\,\ref{f-fits} (top row, first and  third panels).  Given that the
equivalent  widths of  the  Balmer  lines go  through  a maximum  near
$\Teff=13,000$\,K (with the exact value being a function of $\log g$),
$\Teff$ and  $\log g$ determined  by fitting the Balmer  line profiles
are subject to an ambiguity, often referred to as ``hot'' and ``cold''
solutions. This implies that, at  a given equivalent width, the Balmer
lines of  a cold and  massive WD have the  same profile as  the Balmer
lines of  a hot and less  massive WD (see Fig.\,\ref{f-fits},  top row
second  and fourth  panels).   We break  this  degeneracy as  follows.
First, we fit  the entire WD spectrum (continuum plus  lines) with the
same grid  of model  spectra used  for fitting  the Balmer  lines (see
bottom panels of Fig.\,\ref{f-fits}).   We used the 4,000-5,500\,\AA\,
wavelength range covered by the blue  arm of the LSS-GAC spectra. This
is because in  some cases the combined LSS-GAC spectra  are subject to
artifacts  of a  bad  merging  of the  blue  plus  red arm  individual
spectra.  Since the continuum spectrum of  a DA WD is mostly sensitive
to  $\Teff$, the  best-fit value  from the  entire spectrum  indicates
which of the  two solutions is the prefered one.   However, because of
uncertainties in the flux calibration,  the fit to the entire spectrum
can only be used for breaking  the degeneracy between the hot and cold
solutions,  rather  than  for  obtaining a  reliable  set  of  stellar
parameters.  This  in turn implies  that the solution prefered  by the
best-fit  to  the  entire  spectrum  may  be  subject  to  systemmatic
uncertainties.  Thus, in a second step the choice between hot and cold
solution  is  further  guided   by  comparing  the  ultraviolet  GALEX
\citep[Galaxy          Evolution          Explorer;][]{martinetal05-1,
  morrisseyetal05-1} and optical XSTPS-GAC\footnote{The optical fluxes
  are  derived directly  from  the XSTPS-GAC  $g,r,i$ magnitudes.   In
  those   cases   in  which   the   $g$   magnitudes  are   unreliable
  (Fig.\,\ref{f-colcut})  we  either  use  only the  $r,i$  fluxes  as
  guidance or substitute the $g$ magnitudes by those provided by SDSS,
  when available.}  fluxes to the  fluxes predicted from each solution
(see two  examples in Fig.\,\ref{f-solutions}).  This  exercise allows
us to confidently select the correct solution for 75 of our 92 LSS-GAC
DA  white dwarfs.   We use  these 75  LSS-GAC DA  white dwarfs  as the
sample of analysis in this work.

It has been shown that spectroscopic fits that use 1D model atmosphere
spectra such as  those employed in this work  result in systematically
overestimated surface  gravities for  WDs cooler  than $\sim$12,000\,K
\citep{koesteretal09-1, tremblayetal11-1}.   We have thus  applied the
3D  corrections of  \citet{tremblayetal13-1} to  $\Teff$ and  $\log g$
determined  above.  We  then  interpolated the  $\Teff$  and $\log  g$
values     in    the     tables    of     \cite{renedoetal10-1}    and
\citet{althausetal05-1,  althausetal07-1,  althausetal10-2} to  obtain
masses, cooling  ages, absolute $M_{\rm  g}$, $M_{\rm r}$  and $M_{\rm
  i}$ and  bolometric ($M_{\rm  bol}$) magnitudes  for our  WDs. These
cooling sequences  provide absolute magnitudes in  the $UBVRI$ system,
which are  converted into  the $ugriz$ system  using the  equations of
\citet{jordietal06-1}.   Distances  were  finally  obtained  from  the
distance moduli of our targets,  with the XSTPS-GAC $g,r,i$ magnitudes
corrected for extinction using the 3D Galactic extinction map provided
by \citet{chenetal14-1}.

\begin{figure}
\begin{center}
\includegraphics[width=\columnwidth]{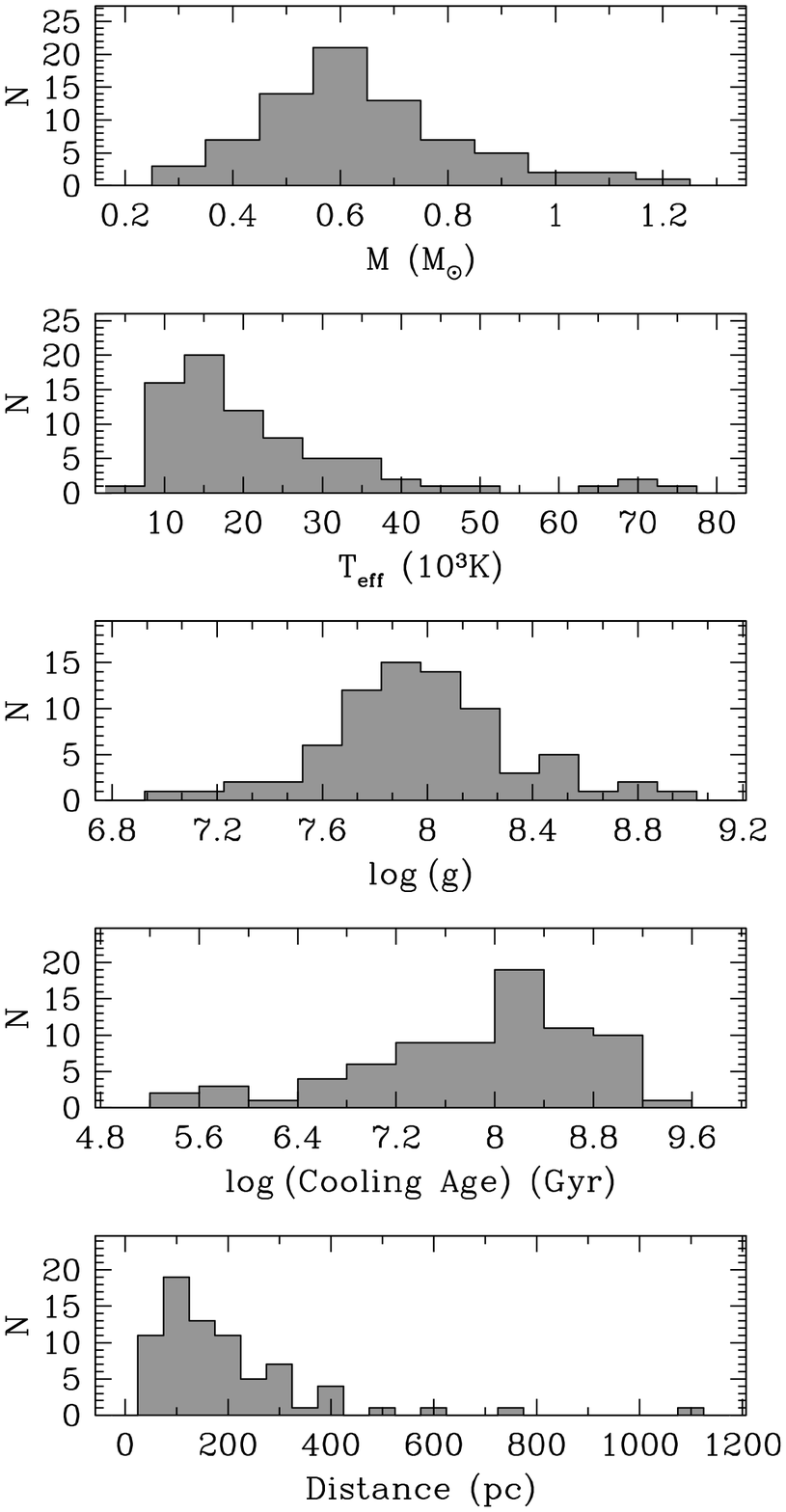}
\caption{\label{f-distr}   From  top   to   bottom:  mass,   effective
  temperature, surface gravity, cooling age and distance distributions
  of the 75 LSS-GAC DA white dwarfs for which we are able to break the
  degeneracy between the hot and  cold solutions obtained from fitting
  the Balmer lines of their spectra.}
\end{center}
\end{figure}

\begin{figure*}
\begin{center}
\includegraphics[angle=-90,width=\textwidth]{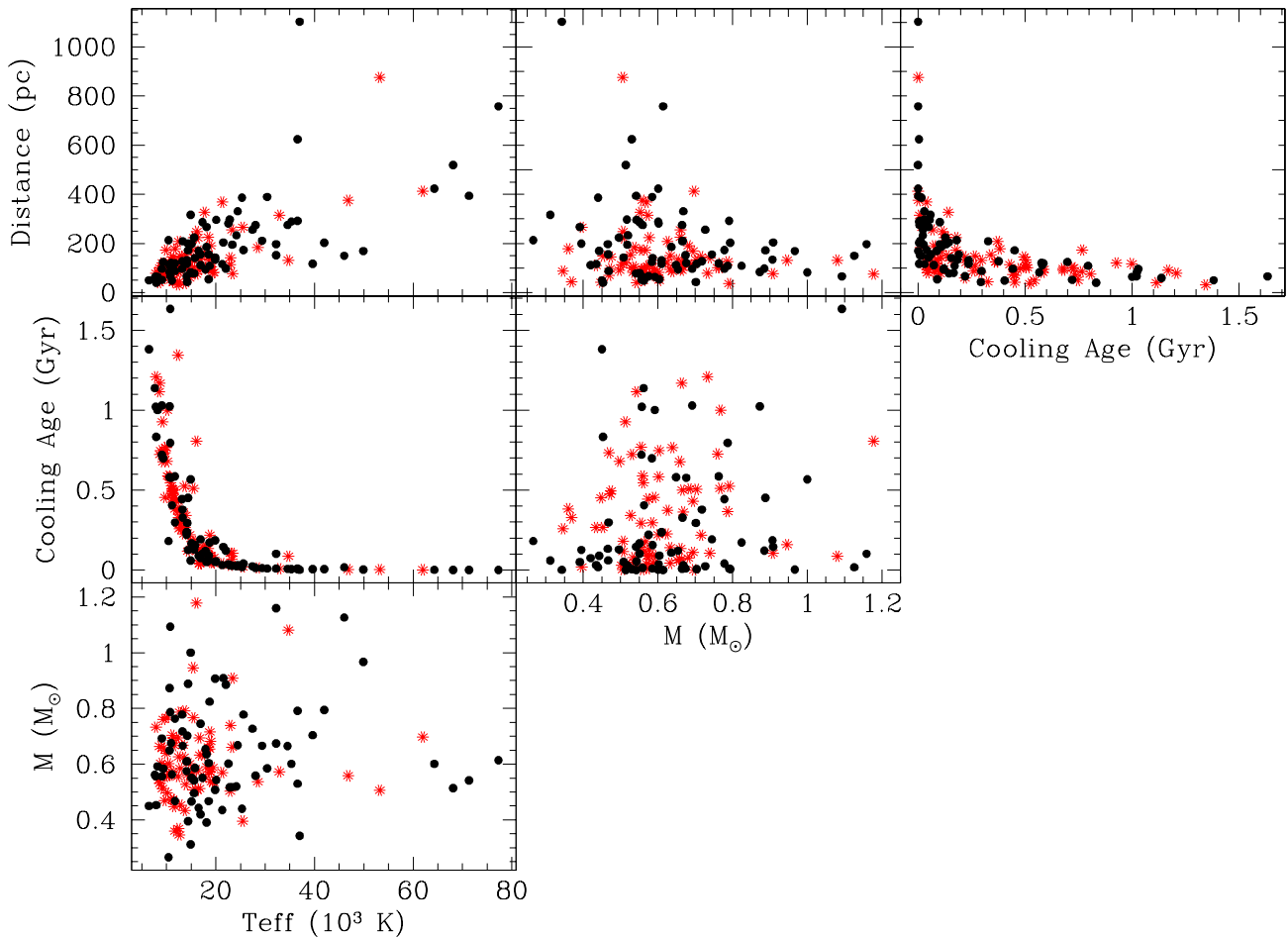}
\caption{\label{f-corr}    Correlations    between    the    effective
  temperatures, masses, distances  and cooling ages of  the 75 LSS-GAC
  DA white dwarfs shown in  Fig.\,\ref{f-distr} (black dots).  We also
  show a typical realization  of the Monte-Carlo simulations described
  in  Section\,\ref{simulations}  (red stars).   As  can  be seen  the
  degree of overlap of the  synthetic sample and the observational one
  is nearly perfect in all six panels.}
\end{center}
\end{figure*}

The  effective temperature,  surface  gravity, mass,  cooling age  and
distance distributions of the 75 DA WDs for which we are able to break
the degeneracy between the hot  and cold solutions obtained by fitting
the   Balmer   lines  of   their   LSS-GAC   spectra  are   shown   in
Fig.\,\ref{f-distr}.  Inspection of this  figure reveals that the vast
majority   of   LSS-GAC   WDs   are   located   at   short   distances
($\sim$50--300\,pc).  The mass and $\log g$ distributions broadly peak
at  0.6\,$M_\odot$ and  $\log g=$\,8  dex respectively,  features that
have     been    observed     previously    in     numerous    studies
\citep[e.g.][]{koesteretal79-1, holbergetal08-1, kepleretal15-1}.  The
effective        temperatures       are        clustered       between
$\sim$\,10,000--15,000\,K, and  the majority of WDs  have cooling ages
between  0.01 and  1\,Gyr.  In  Fig.\,\ref{f-corr} we  show a  grid of
panels displaying the correlations between the effective temperatures,
masses, distances and  cooling ages (black dots) .  We  do not include
the surface  gravity here,  as it  is nearly  equivalent to  mass.  It
becomes clear that the higher  the WD effective temperature the larger
trends to  be the distance,  a clear selection effect  indicating that
cooler WDs  need to be  generally closer to us  to be observed  by the
LSS-GAC.  Also, hotter WDs trend to be further away, as otherwise they
saturate the lower magnitude limit  of the LSS-GAC.  These effects are
also  noticeable  when  inspecting   the  relation  between  mass  and
distance: low-mass  WDs are brighter and  trend to be found  at larger
distances. It  also becomes evident  that, as a simple  consequence of
the WD  cooling, the  cooling ages  decrease for  increasing effective
temperatures.  Because of this there  is also a tight relation between
cooling age and distance, i.e.  shorter cooling ages imply hotter WDs,
therefore further distances. There is also a clear correlation between
the  cooling age  and the  distance, a  quite natural  behavior, since
given  that the  mass distribution  of WDs  has a  narrow peak  around
$0.6\, M_{\odot}$, the cooling ages  and the effective temperature are
very  tightly  correlated (see  the  leftmost  central panel  of  this
figure). Also  shown in this  figure is  a typical realization  of the
Monte         Carlo         simulations        described         below
(Section~\ref{simulations}). It  is quite apparent the  high degree of
overlap  between the  synthetic and  the observed  samples, indicating
that our models reproduce with high fidelity the selection procedures,
and the astronomical properties of the observed population.

From  the above  discussion it  becomes clear  that our  observational
sample  and  corresponding  distributions are  affected  by  selection
effects typical  of those  incorporated by magnitude  limited surveys.
We correct for those observational biases in the next section.

\section{The LSS-GAC DA WD luminosity and mass functions, and the DA WD formation rate}
\label{s-obsfun}

Magnitude limited surveys  such as the LSS-GAC survey  are affected by
selection effects.  Therefore, any parameter distribution that results
from  the  analysis of  a  given  observed  population is  subject  to
observational  biases.   The  $1/V_{\rm   max}$  method  described  in
\citet{schmidt68-1} and  \citet{green80-1} is aimed at  removing these
biases. In  our case this  is done  calculating the maximum  volume in
which each  of our WDs  would have  been detected given  the magnitude
limits of the LSS-GAC survey.  This requires considering the lower and
upper magnitude limits of each of the 16 spectrographs of each LSS-GAC
plate.  For each  spectropgraph, the lower and  upper magnitude limits
define  respectively   the  minimum  ($d_\mathrm{min}$)   and  maximum
($d_\mathrm{max}$)  distances  (and   therefore  minimum  and  maximum
volumes,   $V_\mathrm{min}$  and   $V_\mathrm{max}$)   at  which   the
considered WD would have been detected.  The total maximum volume of a
WD, $V_\mathrm{WD}$,  is the sum  over the individual  maximum volumes
obtained  from   each  spectrograph  --  see   also  \cite{huetal07-1,
  limoges+bergeron10-1}:

\begin{eqnarray}
\label{eq-vol}
V_\mathrm{WD} = V_\mathrm{max}-V_\mathrm{min} = {\sum_{i=1}^{n_{\rm spec}}}\,\frac{\omega_i}{4\pi}\int_{d_\mathrm{min}}^{d_\mathrm{max}}e^{-z/z_0}~4{\pi}r^2dr = \nonumber\\
= -\sum_{i=1}^{n_{\rm spec}} \frac{z_0 \times \omega_i}{\left |\sin{b}\right |}\left [ \left (r^2+\frac{2z_0}{\left |\sin{b}\right |}r+\frac{2z_0^2}{\left |\sin{b}\right |^2}  \right )\! e^{-\frac{r\left |\sin{b}\right |}{z_0}} \right ]_{d_\mathrm{min}}^{d_\mathrm{max}} 
\end{eqnarray}

\noindent where $b$ is the Galactic latitude of the WD, and $\omega_i$
is the  solid angle  in steradians covered  by each  spectrograph (1.2
deg$^2  \times \pi^2/180^2$;  $\sum_{i=1}^{n_{\rm spec}}  \omega_i$ is
the  total area  observed by  the  survey, also  in steradians).   The
factor $e^{-z/z_0  }$ takes into account  the non-uniform distribution
of  stars  in  the   direction  perpendicular  to  the  Galactic  disc
\citep{feltenetal76-1}, where $z=r \times  \sin(b)$ is the distance of
the WD from  the Galactic plane, and $z_0$ is  the scale height, which
is assumed to be  250\,pc \citep{liebertetal05-1, huetal07-1}.  In the
cases where two or more  spectrographs observe the same region of sky,
we consider  the overlapping region with the  largest volume, spanning
between  the smallest  lower  magnitude limit  and  the highest  upper
magnitude limit of the overlapping spectrographs.

Once $V_\mathrm{WD}$ is calculated for  each WD in the observed sample
the space  density of WDs is  simply obtained as $\sum  1/V_{\rm WD} =
0.83 \pm 0.16 \times10^{-3}$ pc$^{-3}$, where the sumation is over all
the  WDs in  the sample\footnote{The  error of  the space  density was
  obtained  artificially  producing  200   versions  of  the  observed
  luminosity function by varying the value of the function in each bin
  with a random value sampled from a poisson distribution proportional
  to the error bar corresponding to that bin.}.  However, it has to be
noted that the space density derived here represents an absolute lower
limit, as we  are able to derive reliable stellar  parameters for only
75 of the  92 DAs in our  sample. That is, we are  considering just 81
per cent of the observed sample in the analysis.  Moreover, the lowest
effective  temperature   value  among  LSS-GAC  DA   white  dwarfs  is
$\sim$6,500\,K.  WDs of lower effective  temperatures are too faint to
be detected by the survey, and  are therefore not accounted for in our
calculation of the space density.  The  space density as a function of
the bolometric magnitude $M_{\rm bol}$,  mass $M_{\rm WD}$ and cooling
age $t_{\rm  c}$ yield the  WD LF, MF  and AF, respectively.   Each of
these functions is analysed in the following sub-sections.

The  $1/V_{\rm  max}$ method  described  above  can  be also  used  to
quantify the completeness of the observed sample, i.e.  the percentage
of  WDs that  are still  missing  because of  selection effects  after
applying the  $1/V_{\rm max}$ method.   This completeness must  not be
taken as the  spectroscopic completeness of the  LSS-GAC sample, which
is  95  per  cent  (see Section\,\ref{s-compl}).   If  the  sample  is
complete,  then the  average  value $\langle  V-V_\mathrm{min}\rangle/
\langle   V_\mathrm{max}-V_\mathrm{min}   \rangle$   should   be   0.5
\citep{green80-1} (where $V$ is the volume of the WD as defined by its
distance, i.e.  the same  as Equation\,(\ref{eq-vol}), but integrating
from 0 to  $d$).  In our case this quantity  is 0.4, which corresponds
to  a completeness  of  80 per  cent.  Of  course,  the above  derived
estimate of  the completeness is  within the context of  the magnitude
limits  of  the  LSS-GAC  survey,   i.e.   it  does  not  account  for
populations  of WDs  that  are  too faint/rare  to  make  it into  the
observed sample.  Moreover, 19 per cent of the observed sample has not
been considered  in the analysis.   If we  were able to  constrain the
stellar parameters of these WDs, then the completeness would increase.

\subsection{The luminosity function}
\label{s-lumfunc}

\begin{figure}
\begin{center}
\includegraphics[angle=-90,width=\columnwidth]{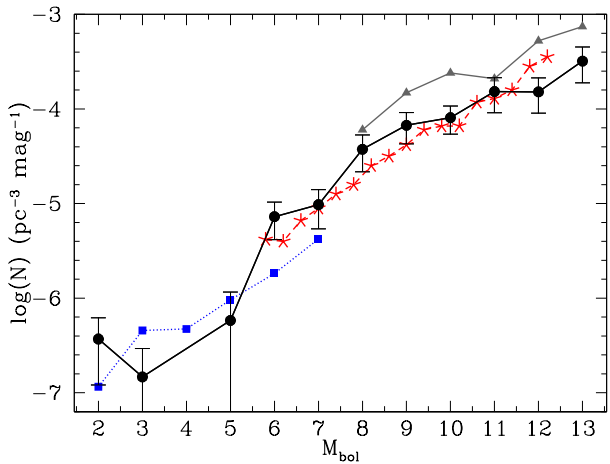}
\caption{\label{f-lf} The LF of LSS-GAC DA white dwarfs and associated
  errors  (solid  black  line  and  solid  dots).   For  the  sake  of
  comparison     we    also     show    the     LFs    obtained     by
  \citet{degennaroetal08-1}   (red  dashed   line   and  red   stars),
  \citet{torresetal14-1}  (blue dotted  line  and  blue squares),  and
  \citet{giammicheleetal12-1} (solid grey line and grey triangles).}
\end{center}
\end{figure}

The LSS-GAC WD LF is  shown in Fig.\,\ref{f-lf}, its associated errors
are calculated  following \citet{boyle89-1}.   For comparison  we show
the  LFs obtained  by \citet{degennaroetal08-1}  for the  SDSS survey,
\citet{torresetal14-1} for hot  DAs in the SDSS  (which supersedes the
LF  of \citealt{krzesinskietal09-1}),  and \citet{giammicheleetal12-1}
for a local and  volume limited sample of WDs.  We  note that two more
LFs are  available from  the SDSS, provided  by \citet{harrisetal06-1}
for  photometric WD  candidates,  and by  \citet{huetal07-1} for  Data
Release  1  spectra of  DA  WDs.   However, we  do  not  show them  in
Fig.\,\ref{f-lf} because (1) the  LF of \citet{harrisetal06-1} assumes
a  constant  $\log$\,g  $=8$\,dex  and very  much  resembles  the  one
provided by  \citet{degennaroetal08-1} when considering DA  WDs of the
same  $\log$\,g  value;  (2)  the  work  by  \citet{degennaroetal08-1}
supersedes the analysis of \citet{huetal07-1}. Moreover, we decide not
to show  the LFs obtained  by \citet{liebertetal05-1} for  the Palomar
Green Survey and \citet{limoges+bergeron10-1}  for the KISO survey, as
they  also resemble  the LF  of \citet{degennaroetal08-1}  but contain
considerably fewer  objects.  Because  of completeness issues,  the LF
derived  by   \citet{rowell+hambly11-1}  for  WD  candidates   in  the
Super-Cosmos survey is  not included neither.  To  avoid clustering of
data in Fig.\,\ref{f-lf} we have also  opted not to show the errors of
all the above mentioned LFs.

Inspection     of      Fig.\,\ref{f-lf}     reveals      that,     for
$M_\mathrm{bol}\geq6$\,mag, the  LF derived  in this  work is  in good
agreement  with the  LF  of  \citet{degennaroetal08-1}.  The  apparent
disagreement   between    our   LF    and   the   one    obtained   by
\citet{giammicheleetal12-1} is likely due to  the fact that the latter
study includes all WDs (not only DAs) in a volume-limited local sample
which  does  not  presumably  suffer  from  completeness  issues,  and
therefore the  space density is higher.   For $M_\mathrm{bol}<6$\,mag,
our LF is also in  broad agreement with that of \citet{torresetal14-1}
for hot DAs.  However, the number of LSS-GAC WDs falling in these bins
is too small for a meaningful  comparison between the two studies.  It
should be  also noted  that the  WD LF  actually continues  to fainter
magnitudes than  those shown  in Fig.\,\ref{f-lf},  however we  do not
display those bins as these objects are too faint to be present in the
LSS-GAC sample.

\subsection{The mass function}

\begin{figure}
\begin{center}
\includegraphics[angle=-90,width=\columnwidth]{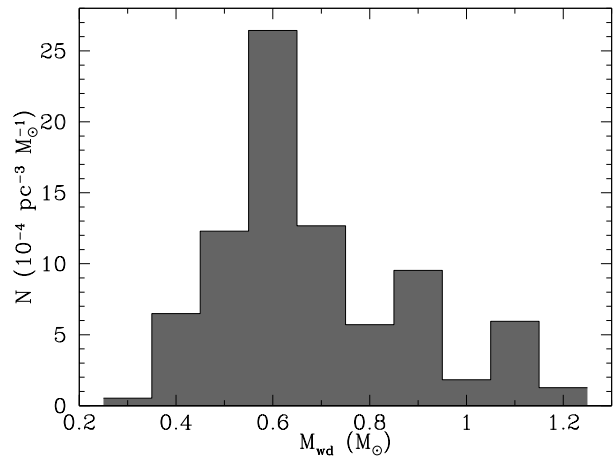}
\caption{\label{f-mf} The MF of LSS-GAC DA white dwarfs.}
\end{center}
\end{figure}

The  MF of  the  LSS-GAC  WDs is  displayed  in Fig.\,\ref{f-mf}.   As
expected, it shows a clear  peak around 0.6 $M_{\odot}$.  A relatively
high percentage  ($\sim$10 per cent) of  low-mass ($<$0.5~$M_{\odot}$)
WDs can also be seen.   Traditionally, the existence of these low-mass
WDs     has     been     attributed     to     binary     interactions
\citep[e.g.][]{liebertetal05-1}, and  indeed it has  been demonstrated
that   the  majority   of  low-mass   WDs  are   formed  in   binaries
\citep{rebassa-mansergasetal11-1,  kilicetal12-1}.  Therefore,  DA WDs
in  those low-mass  bins  are expected  to be  part  of binaries  that
contain  unseen  companions.    Inspection  of  Fig.\,\ref{f-mf}  also
reveals   a   large   fracion   ($\sim$30   per   cent)   of   massive
($\geq$0.8~$M_{\odot}$)  WDs.   A  high-mass  feature  has  also  been
regularly      detected      in      a     number      of      studies
\citep[e.g.][]{liebertetal05-1, kepleretal07-1,  kleinmanetal13-1} and
it  has been  claimed that  it arises  as a  consequence of  WD binary
mergers.   Population synthesis  studies however  do not  predict more
than $\sim$10 per cent of the entire WD population being the result of
binary    mergers     \citep{hanetal94-1,    han98-1,    mengetal08-1,
  toonenetal12-1,   garcia-berroetal12-1}.   Alternatively,   a  large
number  of  high-mass  WDs  in  the  MF  presented  here  may  be  the
consequence  of large  uncertainties in  the mass  determinations (the
mass   errors  of   some  of   our   objects  are   estimated  to   be
$\ga0.1-0.15$\,$M_{\odot}$,  which can  move some  objects across  the
corresponding  mass   bins).   We  will  further   discuss  the  large
percentage   of    high-mass   WDs    identified   in   our    MF   in
Section\,\ref{s-discmass}.

\subsection{The formation rate of DA WDs}
\label{s-formrate}

We now estimate the average DA  WD formation rate following the method
outlined by \citet{huetal07-1}.   If the WD formation  rate is assumed
to be constant  during the last Gyr, then the  slope of the cumulative
AF (see Fig.\,\ref{f-af})  can be considered as  the average formation
rate.  Thus, we simply fit the cumulative AF with a straight line (the
red dashed line in Fig.\,\ref{f-af}),  and identify the slope of $5.42
\pm 0.08 \times10^{-13}$  pc$^{-3}$ yr$^{-1}$ of the fit as  the DA WD
formation  rate.   Inspection  of Fig.\,\ref{f-af}  reveals  that,  as
expected,     the    maximum     of    the     cumulative    AF     is
$0.83\times10^{-3}$\,pc$^{-3}$, i.e.   the total  LSS-GAC DA  WD space
density.

Numerous studies  in the past  two decades have obtained  WD formation
rates. The  most recent analysis \citep{verbeeketal13-1}  results in a
birth  rate of $5.4  \pm 1.5  \times10^{-13}$ pc$^{-3}$  yr$^{-1}$, in
excellent  agreement with  the  value estimated  here.  The  formation
rates   derived   by   \citet[][$2.5-2.7   \times10^{-13}$   pc$^{-3}$
  yr$^{-1}$]{huetal07-1},    \citet[][$6   \times10^{-13}$   pc$^{-3}$
  yr$^{-1}$]{liebertetal05-1},  \citet[][$6  \times10^{-13}$ pc$^{-3}$
  yr$^{-1}$]{holbergetal02-1}  are also  broadly  consistent with  the
value estimated here. For  comparison, earlier studies yield formation
rates    that     are    generally    considerably     higher    (e.g.
\citealt{green80-1}, $20  \pm 10 \times10^{-13}$  pc$^{-3}$ yr$^{-1}$;
\citealt{weidemann91-1},  $23   \times10^{-13}$  pc$^{-3}$  yr$^{-1}$;
\citealt{vennesetal97-1},  $8.5   \pm  1.5  \times10^{-13}$  pc$^{-3}$
yr$^{-1}$).  This  may be a  consequence of the recent  improvement in
quality and size  of WD data sets.  Planetary  nebulae birth rates are
also found  to be higher  in general [\citealt{ishida+weinberger87-1},
  $80 \times10^{-13}$ pc$^{-3}$ yr$^{-1}$; \citealt{phillips02-1}, $21
  \times10^{-13}$ pc$^{-3}$  yr$^{-1}$; \citealt{frew08-1}, 8  $\pm$ 3
  $\times10^{-13}$  pc$^{-3}$ yr$^{-1}$].   The  discrepancy has  been
discussed in detail in \citet{liebertetal05-1}.

\section{The LSS-GAC simulated DA WD luminosity, mass and cumulative age functions}
\label{simulations}

In the previous sections we  presented and characterised the sample of
DA WDs identified  within the data release 1 of  the LSS-GAC.  We also
derived the DA WD space density,  which has been used to construct the
preliminary LF  and MF of LSS-GAC  DA WDs.  Finally, we  estimated the
average DA  WD formation rate from  the DA WD cumulative  AF.  In this
Section we simulate the LSS-GAC DA WD population and take advantage of
the well-defined selection criteria employed  by the LSS-GAC survey to
evaluate the fraction  of simulated WDs that would  have been observed
by the  LSS-GAC survey.  This  will allow  us to directly  compare the
ensemble properties of the observational data sets with the outcome of
the simulations (see Section\,\ref{s-discuss}).

\subsection{The population synthesis code}
\label{s-MC}

\begin{figure}
\begin{center}
\includegraphics[angle=-90,width=\columnwidth]{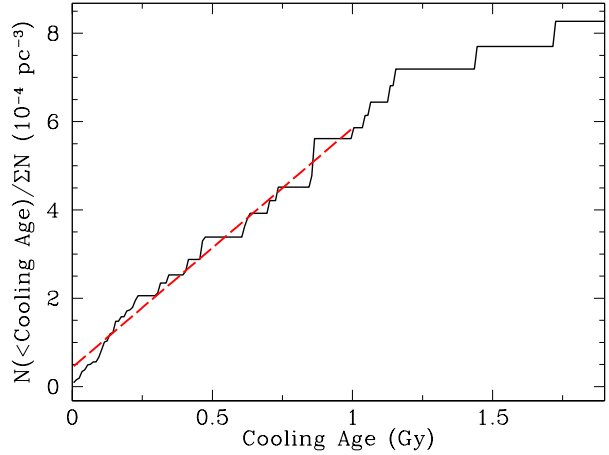}
\caption{\label{f-af}  The  LSS-GAC  DA  white  dwarf  cumulative  age
  function.  The  slope of a  straight line fit  to the last  Gyr (red
  dashed line) gives the average DA WD formation rate.}
\end{center}
\end{figure}

We provide here  a brief description of our Monte  Carlo WD population
synthesis code, and  of the changes done to adapt  it to reproduce the
LSS-GAC.  A  more complete and  detailed description of  the principal
components   of  this   method  can   be  found   in  previous   works
\citep{garcia-berroetal99-1, torresetal02-1, garcia-berroetal04-1}.

Any Monte Carlo code has at its  very core the idea of repeated random
sampling,  i.e.  generating  the statistical  properties of  data from
known   distributions,   that   is   used  to   obtain   the   initial
characteristics (mass, time of birth, initial position, kinematics, as
well as other interesting quantities) of  every star that will come to
form part  of the initial  synthetic population.  One  main ingredient
here  is the  pseudo-random number  generator,  for which  we use  the
algorithm from \cite{james90-1}.  This  produces a uniform probability
density between  $(0,1)$ with a  repetition period of  over $10^{18}$,
more than sufficient for most  practical purposes.  The next important
step is using adequate probability distribution functions for sampling
the  stellar properties.   These  distribution  functions are  crucial
inputs that  define each particular population.   Thus, stellar masses
are   sampled    using   the   initial   mass    function   (IMF)   of
\cite{kroupa01-1}, a  standard choice,  in particular  considering the
alleged universal  character of the IMF  \citep{bastianetal10-1}.  The
moment at which  each star is born ($t_\mathrm{born}$)  is obtained in
accordance with  a star formation  rate (SFR), assumed to  be constant
unless otherwise  specified.  The  position of  each star  is randomly
generated  from  a  double  exponential  distribution  of  a  constant
Galactic  scale height  of $250$\,pc  and a  constant scale  length of
$3.5$\,kpc.   The  velocity distribution  that  we  employ takes  into
account the differential rotation of the Galaxy, the peculiar velocity
of   the   Sun  and   a   scale   height  dependent   dispersion   law
\citep{mihalas81-1}.  Also,  a metallicity  value is assigned  to each
star according to a Gaussian  metallicity distribution as presented in
\cite{casagrandeetal11-1}.

In order to reproduce the LSS-GAC,  stars are only generated in a cone
delimited by $-30^{\circ} \leq b \leq 30^{\circ}$ in Galactic latitude
and  $150^{\circ}  \leq  l  \leq 210^{\circ}$  in  Galactic  longitude
(Section\,\ref{s-gac}), with no restriction  in terms of distance from
the Sun. However, we define a test cone  of up to 200 pc in length, in
which we interactively  examine the density of  generated stellar mass
until we reach a limit density  value for the local stellar population
(e.g.  \citealt{holmberg+flynn00-1}).  We scale this limit in order to
obtain a final restricted WD sample  of the same order as the observed
one.

In  a   next  step  we   set  a  9.5\,Gyr   age  for  the   thin  disk
($t_\mathrm{disk}$)  and  interpolate   the  main  sequence  lifetimes
($t_\mathrm{MS}$)  of  the  generated  stars  using  the  BaSTI  grids
according       to       stellar       mass      and       metallicity
\citep{pietrinfernietal04-1}.         Knowing       $t_\mathrm{disk}$,
$t_\mathrm{MS}$ and $t_\mathrm{born}$, we can easily evaluate which of
those stars have had  time to become WDs.  If that is  the case the WD
cooling     age    is    simply     given    by     $t_\mathrm{c}    =
t_\mathrm{disk}-t_\mathrm{MS}-t_\mathrm{born}$.  Also,  by knowing the
mass  of  the WD  progenitor  we  can compute  the  WD  mass using  an
initial-to-final mass relation (IFMR),  which will be further detailed
in    Section\,\ref{subsec:Models}.    WD    luminosities,   effective
temperatures,  surface  gravities  and  $UBVRI$  and  $M_\mathrm{bol}$
magnitudes are  then obtained by interpolating the  inferred WD masses
and cooling  ages along  the following cooling  tracks: for  WD masses
smaller than 1.1\,$M_{\odot}$ and larger than 0.45\,$M_{\odot}$ we use
the CO sequences  of \cite{renedoetal10-1} and \cite{althausetal10-2},
while for WD masses above the  upper value we employ the ONe tracks of
\cite{althausetal05-1}   and  \citet{althausetal07-1}.    Finally,  we
convert  the $UBVRI$  magnitudes  into the  $ugriz$  system using  the
equations  of  \citet{jordietal06-1},   taking  into  account  the  3D
Galactic  extinction map  of \citet{chenetal14-1}  and  the extinction
coefficients of \citet{yuanetal13-1}.

\begin{table}
\caption{\label{tab:Models} The four models  adopted in this work with
  the aim of reproducing the  observed LSS-GAC DA WD population. Model
  1 is our standard model.}
\begin{center}
\begin{tabular}{cccc}
\hline
\hline
Model & SFR & IFMR & Slope for the\\
      &     &      & massive regime \\
\hline
1 & Constant & \cite{catalanetal08-1}  & 0.10 \\
2 & Constant & \cite{catalanetal08-1}  & 0.06 \\
3 & Constant & \cite{ferrarioetal05-1} & 0.10 \\
4 & Bimodal  & \cite{catalanetal08-1}  & 0.10 \\
\hline
\end{tabular}
\end{center}
\end{table}

Some of  the DA  WD stellar  parameters derived from  the fits  to the
LSS-GAC   spectra    are   subject   to   relatively    large   errors
(Section\,\ref{s-param}).  It  is therefore  necessary to  account for
those uncertainties  in the simulated WD  populations before comparing
the synthetic and observational  data sets.  The effective temperature
errors  of the  observed sample,  which  show a  modest increase  with
increasing temperature, were  fitted by a third  order polynomial such
that  the error  for relatively  cool ($\sim10,000$\,K)  WDs is  about
$300$\,K, increasing to $\sim1,000$\,K for  WDs as hot as $25,000$\,K.
We adopt  this polynomial relation for  deriving effective temperature
errors of  our simulated  WDs. The  observational errors  of $\log\,g$
cluster around $\sim0.2$\,dex,  and we take this value  as the surface
gravity uncertainty  of the  synthetic WDs.   The values  of effective
temperature and surface  gravity for each simulated  WD are re-defined
considering a random value within the  error range defined for the two
quantities.   We  then interpolate  new  values  of mass,  luminosity,
cooling age, and bolometric and absolute magnitudes from the redefined
\Teff\, and $\log g$ values. This  results in, for example, an average
error  in   mass  of  $\sim0.1\,M_{\odot}$  for   our  simulated  WDs.
Photometric errors  are also  taken into  account.  They  are directly
derived  from  the  photometric   uncertainties  associated  with  the
XSTPS-GAC survey \citep{liuetal14-1}.

\subsection{Models}
\label{subsec:Models}

As  presented  in  Section\,\ref{s-obsfun}, the  observational  sample
exhibits several specific features that are clearly visible in the LF,
MF  and cumulative  AF of  the LSS-GAC  WD sample  (Figs.\,\ref{f-lf},
\ref{f-mf} and \ref{f-af}).  We attempt to reproduce those features by
employing the above described population synthesis code.

Given   the  apparent   excess  of   massive  WDs   seen  in   the  MF
(Fig.\,\ref{f-mf}), we  attempt to reproduce this  feature focusing on
three parameters of the simulations that  can affect the final WD mass
distribution, namely the SFR, the IFMR  up to an initial mass of about
6\,$M_{\odot}$ (the mass range of the zero-age main sequence for which
WDs  with CO  cores  are  produced) and  the  slope  of this  (linear)
relationship for  the high-mass  end (WDs with  ONe cores).   We start
with our  fiducial model,  from now  on called model  1, which  uses a
constant   SFR,    the   piecewise    linear   IFMR    introduced   by
\cite{catalanetal08-1} for  the CO WD regime,  and a slope of  0.1 for
the massive regime \citep{ibenetal97-1}.

We then consider three additional models, in which we vary only one of
the above  three parameters with  respect to model  1.  In model  2 we
employ   a   slope   of   0.06   for   the   IFMR   of   massive   WDs
\citep{weidemann05-1}.  The  reason  for  lowering the  slope  of  the
relationship to  this value is to  expand the range of  initial masses
that  can produce  massive WDs  (over 1\,$M_{\odot}$)  in the  hope of
reproducing an excess.  In order to  ensure the continuity of the IFMR
over the entire WD  mass range and to be consistent  with the upper CO
WD mass  limit, we consider that  all stars with masses  between 6 and
11\,$M_{\odot}$  become WDs  of core  masses ranging  from 1.1  to 1.4
$M_{\odot}$, which neatly gives this  slope.  Extending the mass range
up to 1.4 $M_{\odot}$ is probably  wrong, given that a WD that massive
would  most  likely  explode   \citep{ritossaetal99-1},  but  for  the
purposes of  the current  test is an  acceptable assumption.   Model 3
uses the curved  IFMR from \cite{ferrarioetal05-1} for  the CO regime,
which, according to these authors,  results in a better agreement with
the WD mass distribution as compared to when a linear fit is used.  In
model 4  we use the  bimodal SFR  of \cite{rowell13-1}, which  has two
broad  peaks at  around 2  and 7\,Gyr  ago. This  SFR should  favor an
increase in  the number  of massive  WDs during the  last 2  Gyr given
their shorter main  sequence lifetimes.  For each model  we perform 10
individual realizations,  and we compute  the ensemble average  of all
the relevant  quantities. A summary  of the input parameters  used for
each model is given in Table~\ref{tab:Models}.

\subsection{The selection function}
\label{s-selfunc}

Once the synthetic DA WD samples  have been obtained for the different
models  outlined in  the  previous Section,  it  becomes necessary  to
evaluate which of those synthetic WDs  would have been observed by the
LSS-GAC  survey.   Here, we  describe  how  the selection  process  is
performed.

The  first step  is  to  evaluate the  effect  of  the LSS-GAC  target
selection criteria (Section\,\ref{s-gac}).  This is done independently
for each of the 10 realizations  of each model.  For this purpose, the
$g,r,i$  magnitudes of  all WDs  that are  part of  a given  synthetic
population are embedded within the XSTPS-GAC photometric catalogue and
the  selection  criteria  is  then applied  to  the  entire  resulting
population.  The  magnitude limits of  the LSS-GAC survey  are 14\,mag
$\leq  r  \leq$  18.5\,mag (Section\,\ref{s-gac}),  and  within  those
limits the LSS-GAC criteria efficiently  selects 80-85 per cent of the
total simulated WD  population, depending on the  model. This fraction
increases  to 85-90  per cent  if we  consider 14\,mag  $\leq r  \leq$
18\,mag.  The  high success rate  of selecting WDs is  not unexpected,
considering  that  the LSS-GAC  survey  is  specifically developed  to
efficiently target stars  of all colours, including  blue objects such
as WDs (Section\,\ref{s-gac}).

In  a second  step we  evaluate which  of the  simulated WDs  that are
selected  by  the LSS-GAC  criteria  fall  within  the field  of  view
(5\,deg$^2$) of the LSS-GAC  plates actually executed (Table\,A1).  If
this is  the case, an additional  condition is that the  simulated WDs
are   required    to   fulfill    the   magnitude   limits    of   the
plates/spectrographs, otherwise they would not have been observed.  In
practice, we  consider the distances  between the position  defined by
the  right ascension  and declination  of  each simulated  WD and  the
central  positions of  the 16  spectrographs  of the  plate where  the
synthetic  WD  falls  (also  defined by  their  right  ascensions  and
declinations) and  evaluate whether  or not the  $r$ magnitude  of the
synthetic  WD   is  within  the   magnitude  limits  of   the  nearest
spectrograph.  If all those conditions are fulfilled, we then consider
the probability of a given target to be allocated a fibre (some fibres
are used for  sky observations).  This probability is  simply given by
$N_\mathrm{spec}/(N_\mathrm{spec}+N_\mathrm{sky})$,  and is  generally
$\sim$0.9. $N_\mathrm{spec}$ is the  number of target spectra observed
by  the spectrograph,  and $N_\mathrm{sky}$  is the  number of  fibres
allocated for sky observations.

\begin{table}
\caption{\label{t-selfun}  The  synthetic  WD populations  are  passed
  through  consecutive steps  of filtering  that gradually  reduce the
  number of surviving objects.  We show here an example for one of the
  10     performed    realizations     of    our     standard    model
  (Table\,\ref{tab:Models}). In  the last two columns  we indicate the
  percentage of WDs that survive respect  to the previous step and the
  percentage of  WDs that survive  respect to the  initial population,
  respectively.}
\begin{center}
\begin{tabular}{llccc}
\hline
\hline
        & Filter                      &  $N_\mathrm{WD}$ & \% & \% \\
\hline
        & Initial sample                  &  3,874 &     &    \\
        & Initial sample within magnitude &  2,132 &     &    \\
        & limits of the LSS-GAC           &        &     &    \\
 Step 1 & Selection Criteria              &  1,748 & 82  & 82 \\
 Step 2 & GAC plates + fibre allocation   &    192 & 11  & 9  \\
 Step 3 &  S/N $\geq$5                    &    102 & 53  & 5  \\
 Step 4 & Completeness + spectral fit     &     77 & 75  & 3.5\\
\hline
\end{tabular}
\end{center}
\end{table}

\begin{figure*}
\begin{center}
\includegraphics[angle=-90,width=\textwidth]{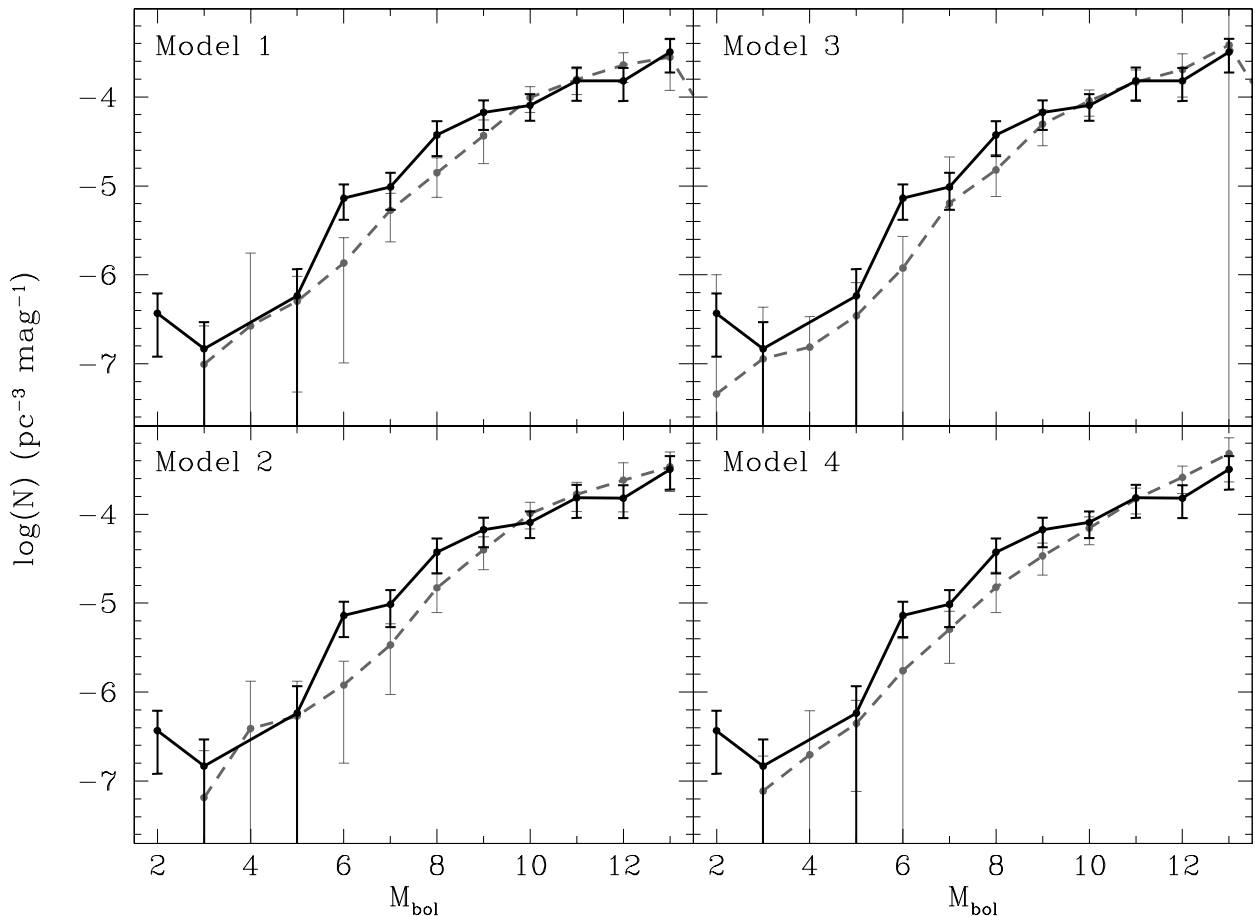}
\caption{\label{f-LF} Simulated DA WD LFs (grey solid lines) resulting
  from  the  models  considered in  Section\,\ref{subsec:Models}  (see
  Table\,\ref{tab:Models}).  The observed LSS-GAC WD  LF is shown as a
  black  solid line.}
\end{center}
\end{figure*}

If the  synthetic WDs survive  all the previously  explained filtering
process  we consider  the  LSS-GAC survey  would  have observed  them.
Therefore,  in a  third  step  we consider  the  probability for  each
simulated WD to have  a LAMOST spectrum of S/N ratio  $\geq$ 5 in both
the blue and red arms.  For  each synthetic WD we calculate the fluxes
from their associated $g$ and $r$ magnitudes, add and subtract a 5 per
cent of  flux in each  case and  calculate the magnitudes  that result
from this exercise  ($g_+$, $g_-$ and $r_+$, $r_-$;  where the sufixes
$+$ and $-$ indicate that we have  added and subtracted the 5 per cent
of the corresponding flux).  We  then consider all targets observed by
the  respective   spectrograph  (i.e.   the  spectrograph   where  the
simulated WD falls)  having $g_- < g <  g_+$ and $r_- < r  < r_+$, and
calculate the  median S/N ratio  of their  LSS-GAC spectra in  the two
bands.   If  no  observed  spectra  are  found  satisfying  the  above
magnitude ranges, or  if one of the median S/N  ratios is smaller than
5, the synthetic WD is then excluded from the analysis.  This exercise
takes  into account  nigh-to-night variations  of S/N  ratio that  may
arise e.g.  from varying observing conditions, as the S/N is evaluated
specifically for objects observed during  the same night with the same
plate/spectrograph, and of  similar magnitudes as the  simulated WD of
concern.

Finally,  in a  fourth step  we  take into  account the  spectroscopic
completeness of  the observed sample  (the fraction of LSS-GAC  DA WDs
that we have identified among all DA WDs observed) as well as consider
the fact  that we are not  able to obtain reliable  stellar parameters
for  19  per  cent  of  the observed  sample.   We  have  estimated  a
spectroscopic  completeness of  95 per  cent (Section\,\ref{s-compl}).
Therefore, we  randomly exclude 5 per  cent of all synthetic  WDs that
passed the  previous filters.   After this  correction, we  proceed by
randomly excluding 19 per cent of the surviving systems.

In order to minimize the effects  of the random exclusion of synthetic
WDs,  we repeat  steps two  to four  20 times  per model  realization.
Given     that    each     of    the     four    models     considered
(Table\,\ref{tab:Models})  counts  10   realizations,  we  obtain  200
different final  synthetic populations for  each model. The  number of
simulated WDs that  pass the entire selection  process described above
vary slightly from model to model (and realization to realization) and
yields synthetic samples  of 65--85 objects, similar to  the number of
WDs in the observed  sample, 75 DA WDs.  An example  of how the number
of synthetic WDs  gradually decreases as they are  passed through each
of   the   filters   of   our    selection   process   is   shown   in
Table\,\ref{t-selfun}.   In   a  final   step  we   use  bootstrapping
techniques to produce synthetic samples  of the same number of objects
as the observed one.

\begin{figure*}
\begin{center}
\includegraphics[angle=-90,width=\textwidth]{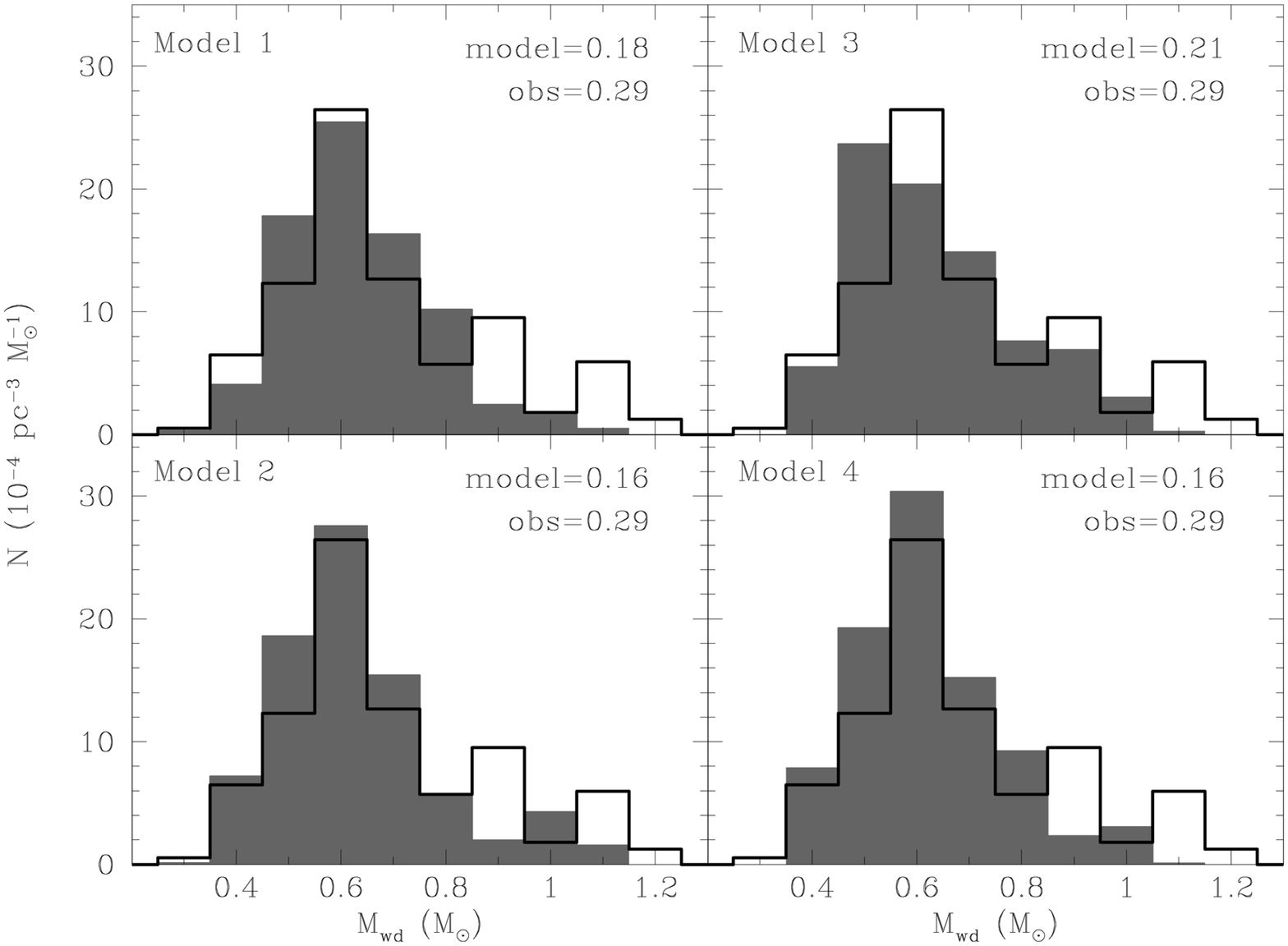}
\caption{\label{f-MF} Same  as Fig.\,\ref{f-LF} but for  the MFs.  The
  fraction of massive WDs  ($>0.8$\,$M_{\odot}$) for both the observed
  and the  simulated samples  are provided  in the  top right  of each
  panel.}
\end{center}
\end{figure*}

The final LF,  MF and cumulative AF  for each model are  the result of
averaging   200  individual  functions  derived   from each   of   the
independent  realizations.   These  are  shown  in  Figs.\,\ref{f-LF},
\ref{f-MF} and \ref{f-AF}, respectively, where we also include the LF,
MF  and   cumulative AF  derived  from  the  observational  sample.  A    
comparison of  the synthetic and    observed  functions  is  presented
and discussed  in  detail  in  the following Section.

\section{Discussion}
\label{s-discuss}

In this section we compare the LFs, MFs and cumulative AFs (i.e. birth
rates) that  result from  our numerical  simulations to  those derived
observationally.   Before comparing  in detail  the simulated  and the
observed distributions,  we compare  the WD populations  obtained from
each of the  models employed here.  We also estimate  the number of DA
WDs that the LSS-GAC will eventually observe.

\subsection{The  final expected number of LSS-GAC DA white dwarfs}

The LSS-GAC  selection criteria (Section\,\ref{s-gac}) applied  to our
simulated  WD  populations results  in  $\sim$80-85  per cent  of  the
synthetic DA  WDs falling  within the magnitude  limits of  the survey
being selected for observations (Section\,\ref{s-selfunc}).  The total
number  of WDs  generated by  each model  oscilates between  3,800 and
3,900,  which reduces  to 2,130--2,160  if we  consider the  magnitude
limits   of   the   LSS-GAC.    This   implies   that,   on   average,
$\sim$1,700-1,850 DA WDs  could be potentially observed at  the end of
the survey, depending on the assumed model.

We have  shown that $\sim$50  per cent of  our synthetic DA  WDs would
have    S/N$\ge$5     if    observed     by    the     LSS-GAC    (see
Table\,\ref{t-selfun}). This percentage is  expected to increase up to
$\sim$2/3 for  the data  release 2 (and  further releases)  of LSS-GAC
spectra (private communication). The number of LSS-GAC DA white dwarfs
expected to have spectra of S/N$\ge$5 at the end of the survey is thus
$N_\mathrm{DA}=$(1,700--1,850$\,\,\,  - \,\,\,  N_\mathrm{obs}) \times
2/3 \simeq$ 1,070--1,170, where  $N_\mathrm{obs}=92$, i.e.  the number
of currently observed DA  WDs.  Considering spectroscopic completeness
and     spectral     fitting    effects     (Section\,\ref{s-selfunc},
Table\,\ref{t-selfun}), which exclude  $\sim25$ per cent of  the DA WD
spectra with  S/N$\ge$5, the final  number of LSS-GAC DA  white dwarfs
with  available and  reliable stellar  parameters  at the  end of  the
survey  is expected  to be  $\simeq$800--875, i.e.   approximately one
order of magnitude higher than  the current number of observed LSS-GAC
DA white dwarfs with reliable stellar parameters.

\subsection{Effects of observational uncertainties}
\label{s-compmod}

We have employed  four different models to simulate  the WD population
in the Galactic  anti-center with the aim of constraining  what set of
assumptions (SFR, IFMR,\ldots) fits  better the observational data. As
expected,  the  intrinsic  properties  of  the  simulated  populations
differed from model  to model.  However, these  properties are altered
when    the     observational    uncertainties     are    incorporated
(Section\,\ref{s-MC}).  For example,  the simulated mass distributions
become broader and  lose detail, and more importantly,  peak at larger
values   (e.g.    the  median   of   the   distribution  shifts   from
0.57\,$M_{\odot}$  to  0.60\,$M_{\odot}$  for Model  2).   Hence,  the
incorporation of observational uncertainties results in less prominent
differences between the synthetic WD parameter distributions.

This  effect is  enhanced  when  we take  into  account the  selection
biases.  In  order to  illustrate these effects  together, we  show in
Fig.\,\ref{f-corr}    the   correlations    between   the    effective
temperatures,  masses, cooling  ages and  distances for  the synthetic
population (red  stars) and  compare them to  those obtained  from the
observational sample (black  dots).  For the seek of  clarity we chose
one typical realization of our  Model 1, although very similar results
are  obtained  for the  other  realizations  and models.   It  becomes
obvious that  the model  reproduces well  the observational  data, and
that  the correlations  between the  considered parameters  follow the
same pattern as the observational one (Section\,\ref{s-param}).

\subsection{The luminosity function}

Fig.\,\ref{f-LF} shows the LFs derived  from our simulated samples, as
well as that  deduced from the observational  data.  The uncertainties
in the  simulated functions were  derived in the  same way as  for the
observed one  (Section\,\ref{s-lumfunc}).  The space  density obtained
for  models  1,   2,  3  and  4,   are  $0.96\pm0.19$,  $0.98\pm0.21$,
$1.16\pm0.20$ and $1.06\pm0.20  \times10^{-3}$ pc$^{-3}$ respectively.
Although  these values  are  slightly higher  than  the space  density
derived from our observations ($0.83\pm 0.16 \times10^{-3}$ pc$^{-3}$,
Section\,\ref{s-obsfun}), they perfectly match  within the error bars.
It is  evident that  there is  an overall  good agreement  (within the
error  bars) between  the simulated  and the  observed LFs,  except at
$M_\mathrm{bol}$  2  and 6\,mag,  where  the  observed LF  predicts  a
considerably higher space density.  It  has to be noted, however, that
the number  of targets falling within  bins of $M_\mathrm{bol}<7$\,mag
is  small (18  per cent  of the  total observed  sample).  Hence,  the
observed LF  in those high  luminosity bins  is subject to  low number
statistics  and the  apparent increase  of  the observed  LF at  those
specific bins should be taken with some caution.  A further inspection
of  Fig.\,\ref{f-LF} reveals  that, because  of the  reasons explained
above (Section\,\ref{s-compmod}),  no model  seems to have  an obvious
advantage in reproducing the observational data.

\subsection{The mass function}
\label{s-discmass}

After applying  the LSS-GAC target  selection criteria and  the target
selection  process   described  in  Section\,\ref{s-selfunc}   to  our
simulated populations, the  MFs yielded by all  simulations are rather
similar (Fig.\,\ref{f-MF}).   In addition,  synthetic (single)  WDs of
masses as low  as 0.35\,$M_{\odot}$ are now possible  as a consequence
of  incorporating  observational  uncertainties.  This  effect  partly
explains     the      apparent     over-abundance      of     low-mass
($\la0.45$\,$M_{\odot}$) WDs in  the observed MF (black  solid line in
Fig.\,\ref{f-MF}).   Alternatively,  a   relative  large  fraction  of
low-mass WDs in the observed sample could be the result of binary star
evolution   \citep{rebassa-mansergasetal11-1}.   The   companions  are
likely to  be cooler  and/or more massive  WDs, or  low-mass late-type
main sequence  stars, although other  exotic companions such  as brown
dwarfs cannot be ruled out.

\begin{figure}
\begin{center}
\includegraphics[angle=-90,width=\columnwidth]{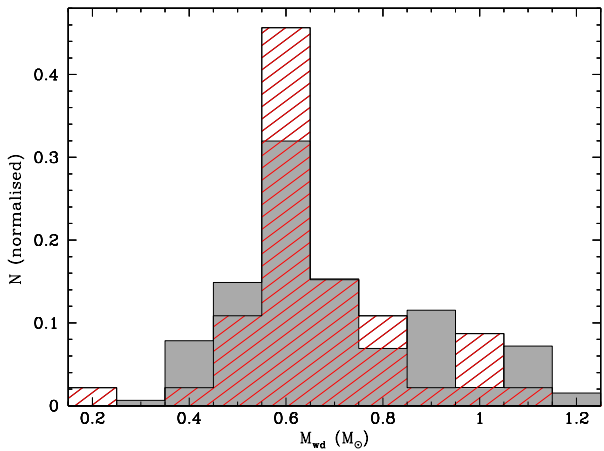}
\caption{\label{f-masscomp}  Normalised  LSS-GAC  DA  white  dwarf  MF
  (gray) and normalised mass distribution of DA WDs from the local and
  volume-limited sample of \citet{giammicheleetal12-1} (red dashed).}
\end{center}
\end{figure}

It  is also  clear  that  none of  our  models  manages to  completely
reproduce the observed behaviour at high mass bins, i.e.  the fraction
of massive WDs ($>0.8$\,$M_{\odot}$) relative to those of typical mass
(0.6\,$M_{\odot}$)  is  higher in  the  observed  sample.  We  discuss
possible scenarios leading to this feature below.

\subsubsection{The initial-to-final mass relation}

The  currently available  IFMRs have  been derived  from observational
data that exhibit  large scatter in the  initial-to-final mass diagram
\citep[see for example Fig.\,1  of][]{catalanetal08-1}.  In one of our
models we  have investigated the  effect of  varying the slope  of the
IFMR   for   producing   a   wider   range   of   massive   WDs   (see
Section\,\ref{subsec:Models}  and  Table\,\ref{tab:Models}).  We  have
also   explored    the   effect    of   employing   a    curved   IFMR
\citep{ferrarioetal05-1}.  To  further investigate  the impact  of the
large scatter  in the initial-to-final  mass diagram to  the simulated
MF, two  additional models (models  5 and  6) are developed  that take
into account the error bars  of the IFMR of \citet{catalanetal08-1} so
that the IFMR  is virtually moved ``up'' in one  model and ``down'' in
the other model.  The remaining free  parameters of models 5 and 6 are
the  same as  for our  standard model  (Table\,\ref{tab:Models}).  The
results show that the MFs obtained from these two models do not differ
significantly from  those shown in Fig.\,\ref{f-MF}  and therefore are
not able of  reproducing the high-mass excess present  in our observed
MF.

\subsubsection{S/N ratio and 3D model atmosphere correction effects}

Two  additional  plausible  explanations  for the  large  fraction  of
massive WDs  observed are effects of  limited S/N ratios and  3D model
atmosphere corrections.

The LSS-GAC spectra  considered in this work have a  minimum S/N ratio
of 5.  It is therefore  possible that some systematic uncertainties in
the WD  stellar parameters result  as a consequence of  the relatively
low S/N ratio of some WD spectra.  This may lead to the masses of some
WDs being overestimated.  In order  to investigate this possibility we
re-derive the observed MF excluding all  systems with spectra of a S/N
ratio below  8. This  leaves us  with 52  DA WDs.   We decided  not to
increase  the S/N  threshold to  higher values  because otherwise  the
number of massive  (fainter and with systematically  lower S/N ratios)
WDs that would survive the cut would be severely reduced.  The MF that
results from  this exercise does  not differ significant from  the one
obtained using the full sample, and  displays as well a large fraction
of massive WDs.   Therefore, S/N ratio effects are unlikely  to be the
cause of the excess of massive WDs observed.

\begin{figure*}
\begin{center}
\includegraphics[angle=-90,width=\textwidth]{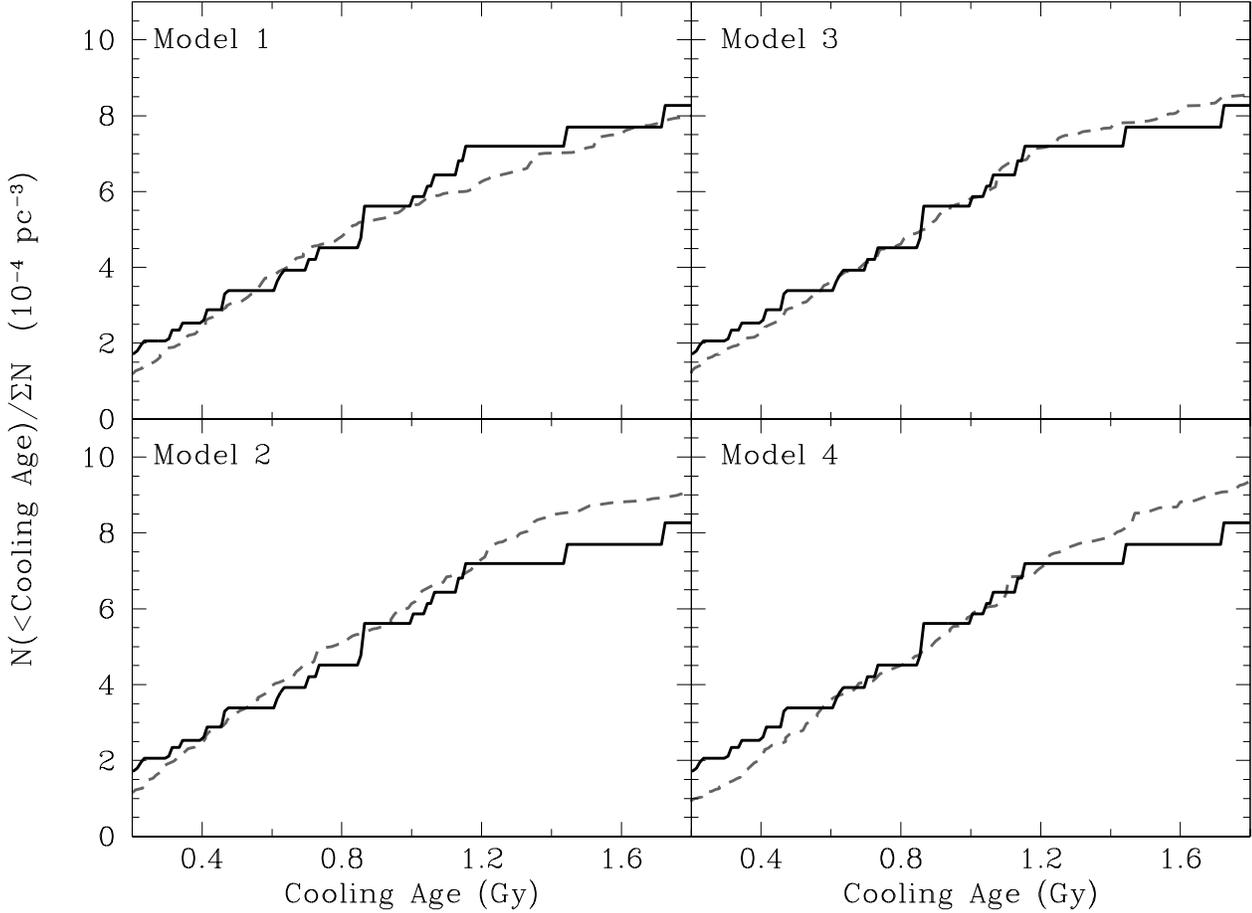}
\caption{\label{f-AF} Same as Fig.\,\ref{f-LF}  but for the cumulative
  AFs.}
\end{center}
\end{figure*}

The DA WD sample analysed in this  work includes cool WDs for which we
have  applied the  3D model  atmosphere corrections  to their  stellar
parameters  deduced  from  1D   model  atmosphere  fitting:  effective
temperature, surface  gravity, and  hence mass.  If  those corrections
are somehow incorrect,  they may lead to an  apparent overabundance of
massive  WDs.  To  explore  this  possibility  we  re-derived  the  MF
excluding all  WDs in our  sample with an effective  temperature below
13,000\,K.  This results in a sub-sample of 58 DA WDs.  The MF deduced
from this sub-sample  again presents a clear  overabundance of massive
WDs.   We therefore  find that  the  overabundance of  massive WDs  is
unlikely  caused by  the  possible  effects related  to  the 3D  model
atmosphere corrections.

\subsubsection{Effective temperature and surface gravity error effects}

The effective temperature and  surface gravity errors obtained fitting
the Balmer lines are not  independent. The reason for this correlation
is that the strength of the  Balmer lines is largely determined by the
ionization balance. That is, if the equivalent width of e.g.  H$\beta$
is fixed,  a higher assumed  effective temperature will need  a higher
surface gravity, as the higher  pressure is needed to compensate.  The
line shape is  the second order effect, which determines  where on the
correlation line the best solution is.  In order to ivestigate whether
or not this  effect may explain the excess of  high-mass WDs observed,
our simulations should have taken into  account not only the errors in
these quantites  (see Sections\,\ref{s-selfunc}  and \ref{s-compmod}),
but also their correlation.

However,  the strength  of this  correlation not  only depends  on the
effective temperature range,  but also the first order  change of line
strength disappears,  and with  it the  correlation, near  the maximum
strength of the line. Moreover, the correlation is not always apparent
for  small  errors. Hence,  quantifying  the  correlation between  the
effective  temperature  and  surface   gravity  errors  is  a  notable
endeaveour, which is beyond the scope  of this paper. Thus, whether or
not such a correlation may explain  the apparent excess of massive WDs
remains an open question.

\subsubsection{WD+WD mergers}

An exciting  possible explanation for  the excess of high-mass  WDs in
the observed MF is that a relatively large fraction of those stars are
the  result  of  mergers  of two  low-mass  WDs  \citep{marshetal97-1,
  vennes99-1}.   Although  no   population  synthesis  study  hitherto
predicts such a  large fraction of high-mass WDs as  the outcome of WD
mergers     \citep[e.g.][]{hanetal94-1,     han98-1,     mengetal08-1,
  toonenetal12-1},  this scenario  has  been adopted  in  some of  the
previous    observational   studies    \citep[e.g.][]{liebertetal05-1,
  giammicheleetal12-1}.   To further  investigate  this hypothesis  we
compare in  Fig.\,\ref{f-masscomp} the normalised MF  obtained in this
work with  the normalised mass  distribution of  DA WDs in  the local,
volume-limited  sample  of  \citet{giammicheleetal12-1}.   In  becomes
obvious that  the peak at  0.6\,$M_{\odot}$ is less pronounced  in our
normalised  MF, an  effect  likely related  to the  fact  that we  are
subject  to  larger  observational  uncertainties  which  broaden  the
distribution.   Interestingly,  whilst  the   high-mass  peak  in  the
normalised mass  distribution of \citet{giammicheleetal12-1}  is found
at 1\,$M_{\odot}$, our  normalised MF shows two apparent  peaks at the
0.9 and  1.1\,$M_{\odot}$ bins and  reflects a scarcity of  systems at
the 1\,$M_{\odot}$  bin.  Although this  discrepancy is likely  due to
our larger uncertainties, which are sufficient to shift objects across
bins, both studies favour the hypothesis that an excess of massive WDs
seems to exist.

If this excess of massive WDs may  arise as a consequence of WD binary
mergers, then the  high merger rate implied by the  observed excess of
massive  WDs  may   indicate  a  much  more  important   role  of  the
double-degenerate  channel for  the production  of Type  Ia supernovae
\citep{wang+han12-1, toonenetal12-1}.   Analysing the  expected merger
rates  of WDs  based on  our observations  and comparing  them to  the
observed rates  of Type Ia  supernovae thus  seems to be  a worthwhile
exercise,  and we  will  pursue this  elsewhere.   High-mass WDs  that
result    from     mergers    are    expected    to     be    magnetic
\citep{garcia-berroetal12-1}, therefore we may expect to find signs of
magnetic  fields  in  our  massive  WDs that  should  help  test  this
hypothesis.

The large WD merger rate suggested  by the current work indicates that
an even  larger number of  close WD binaries  may exist in  the Galaxy
that have not yet merged.  Those close binaries could be a main source
of gravitational waves to be detected by future facilities such as the
space interferometer  eLISA \citep{nelemans13-1}.   Therefore, indirect
support in favour of the merger  scenario may be obtained by analysing
the population of close double WDs that eLISA will discover.

\subsection{The average DA WD formation rate}

The  cumulative  AFs  derived   from  our  simulated  populations  are
illustrated in  Fig.\,\ref{f-AF}, where the observed  cumulative AF is
also  displayed.   There is  an  overall  good agreement  between  our
simulations  and the  osbervations  for coolinjg  ages  up to  1\,Gyr,
except  perhaps  for  our  model  4,  which  seems  to  systematically
overestimate the space density for  cooling age bins $<$0.4\,Gyr (note
that in model  4 we are employing a bimodal  star formation rate). For
cooling ages larger  than 1\,Gyr the discrepancies  between the models
and  the observations  arise  due  to the  scarcity  of  WDs at  those
specific cooling ages.

Fitting  the  simulated  cumulative  AFs with  a  straight  line  (see
Section\,\ref{s-formrate}) we derive average  DA WD formation rates of
6.04$\pm$0.05,   6.42$\pm$0.05,    5.85$\pm$0.02   and   5.97$\pm$0.04
$\times10^{-13}$  pc$^{-3}$  yr$^{-1}$  for  models  1,  2,  3  and  4
respectively.   These values  agree  with the  average formation  rate
derived  from  our observations  within  the  errors ($5.42  \pm  0.08
\times10^{-13}$   pc$^{-3}$   yr$^{-1}$,   Section\,\ref{s-formrate}).
Because of the observational uncertainties (Section\,\ref{s-compmod}),
we  find  that  no  model  seems  to  have  an  obvious  advantage  in
reproducing the observational data.

\section{Summary and conclusions}
\label{s-concl}

The  recently initiated  LAMOST Spectroscopic  Survey of  the Galactic
anti-center,   the   LSS-GAC,   selects  targets   for   spectroscopic
observations  following  a  well-defined criteria.   This  significant
advantage  over  previous   surveys  has  allowed  us   to  present  a
well-characterised    magnitude-limited   sample    of   92    LSS-GAC
hydrogen-rich  (DA)  white  dwarfs  from  the  data  release  1.   Our
catalogue  is expected  to be  $\sim95$  per cent  complete.  We  have
determined   the  stellar   parameters  (surface   gravity,  effective
temperature  and  mass),  absolute   and  bolometric  magnitudes,  and
distances  of  75  DA  white   dwarfs.   Taking  into  account  volume
corrections  we have  derived an  absolute lower  limit for  the space
density of  DA white  dwarfs of  0.83$\pm0.16\times10^{-3}$ pc$^{-3}$.
We  have also  obtained preliminary  observed LSS-GAC  DA white  dwarf
luminosity,  mass  and  cumulative   age  functions.   The  luminosity
function resembles those found in previous observational studies.  The
mass function reveals an excess of massive white dwarfs.  Finally, the
DA white dwarf formation rate derived from the cumulative age function
is  5.42$\pm0.08   \times$10$^{-13}$  pc$^{-3}$  yr$^{-1}$,   in  good
agreement with other recent studies.

We  have simulated  the  DA  white dwarf  population  in the  Galactic
anti-center  using  an  existing  Monte  Carlo  code  adapted  to  the
characteristics  of  the LSS-GAC.   For  this  purpose, and  specially
aiming at reproducing the observed  excess of massive white dwarfs, we
have  employed four  different  models.  All  those  models take  into
account the observational uncertainties,  both spectroscopic (i.e., we
incorporate errors  in the stellar  parameters of our  simulated white
dwarfs based  on the observational  errors) and photometric.   We find
that  the  LSS-GAC  criteria  selects  $\sim$80-85  per  cent  of  all
simulated white  dwarfs with  $14\leq r \leq18.5$\,mag  (the magnitude
limits of  the survey) in  each model, thus providing  robust evidence
for the  high efficiency of  LSS-GAC in targeting white  dwarfs.  Once
the observational  uncertainties have been  taken into account  in our
simulations, the  distribution of  stellar parameters are  similar for
all  models.  We  find that  all  our simulations  reproduce well  the
observed luminosity function, however no particular model seems to fit
better the data.  

None of our considered models is able to reproduce the observed excess
of   massive  DA   white  dwarfs.    We  have   investigated  possible
explanations for this feature and  concluded that a plausible scenario
is that a sizable fraction of  those massive white dwarfs are products
of mergers  of two initially lower-mass  white dwarfs. If that  is the
case, then the  white dwarf merger rate in our  Galaxy is considerably
higher than  currently assumed.  This may  have important implications
for the  production of  Type Ia  supernovae via  the double-degenerate
channel.

Finally, it is important as well  to emphasise that although our study
represents an important step  forward towards unveiling the underlying
population of DA  white dwarfs in the Galaxy, the  size of the LSS-GAC
sample is small,  and that the stellar parameters we  derived for some
objects are  subject to  relatively large  uncertainties.  Forthcoming
LSS-GAC data releases are expected to  increase the number of DA white
dwarfs by  one order of  magnitude.  In  addition, the quality  of the
LAMOST spectra  will improve, which  will reduce the  uncertainties in
the stellar  parameter determinations.   We will hence  derive updated
luminosity and  mass functions and  DA white dwarf formation  rates at
the end of the survey.

\section*{Acknowledgments}

We  thank  the anonymous  referee  for  the relevant  suggestions  and
comments that  helped improving  the paper.  ARM  acknowledges helpful
discussions with X.-B.  Wu, Z.  Han,  X.  Chen, X.  Meng, M.  Zemp, S.
Justham,  T.R.  Marsh  and  G.  Herczeg.   ARM acknowledges  financial
support  from the  Postdoctoral  Science Foundation  of China  (grants
2013M530470   and  2014T70010)   and  from   the  Research   Fund  for
International  Young  Scientists  by   the  National  Natural  Science
Foundation of  China (grant  11350110496).  RC  acknowledges financial
support  from the  FPI grant  BES-2012-053448 and  the mobility  grant
EEBB-I-14-08602. XWL,  HBY, MSX and  YH are supported by  National Key
Basic Research  Program of China  2014CB845700. The work of  EG-B, ST,
and RC was partially funded by  MICINN grant AYA2011-23102, and by the
European Union FEDER funds.

Guoshoujing  Telescope   (the  Large   Sky  Area   Multi-Object  Fiber
Spectroscopic  Telescope,  LAMOST)  is  a  National  Major  Scientific
Project which is  built by the Chinese Academy of  Sciences, funded by
the  National  Development and  Reform  Commission,  and operated  and
managed by the National Astronomical Observatories, Chinese Academy of
Sciences.

\appendix

\section{Tables}

\begin{table*}
\caption{\label{t-plates} LSS-GAC  plates, including  associated right
  ascensions  and declinations,  used  for observations.  The date  of
  observations  are   also  included.}   
\setlength{\tabcolsep}{0.9ex}
\centering
\begin{small}
\begin{tabular}{cccccccccccc}
\hline
\hline
Date & Plate & RA & DEC & Date & Plate & RA & DEC  \\
     &    & [deg] & [deg]    & & & [deg] & [deg] \\
\hline
 2011-10-03  &            PA09B\_keda1 &    63.79331 &   29.90206 &    2012-10-12 &               GAC090N33B1 &   90.06008 &   33.13693  \\ 
 2011-10-03  &            PA09B\_keda2 &    63.79331 &   29.90206 &    2012-10-13 &               GAC072N32B1 &   72.32949 &   32.58819  \\ 
 2011-10-03  &            PA09M\_keda1 &    63.79331 &   29.90206 &    2012-10-13 &               GAC072N32M1 &   72.32949 &   32.58819  \\ 
 2011-10-03  &            PA09M\_keda2 &    63.79331 &   29.90206 &    2012-10-17 &               GAC083N27B1 &   83.98125 &   27.66235  \\ 
 2011-10-05  &            PA09F\_keda1 &    63.79331 &   29.90206 &    2012-10-17 &               GAC083N27B2 &   83.98125 &   27.66235  \\ 
 2011-10-05  &            PA09M\_keda1 &    63.79331 &   29.90206 &    2012-10-18 &               GAC085N33B2 &   85.74273 &   33.31455  \\ 
 2011-10-05  &            PA09M\_keda2 &    63.79331 &   29.90206 &    2012-10-19 &               GAC080N32B1 &   80.00299 &   32.78560  \\ 
 2011-10-21  &                    PA1B &    45.87566 &   28.26991 &    2012-10-19 &               GAC080N32F1 &   80.00299 &   32.78560  \\ 
 2011-10-21  &                    PB8B &    91.59353 &   29.51241 &    2012-10-19 &               GAC080N32M1 &   80.00299 &   32.78560  \\ 
 2011-10-21  &                    PB8M &    91.59353 &   29.51241 &    2012-10-24 &               GAC051N24B1 &   51.07698 &   24.72406  \\ 
 2011-10-28  &          GAC\_060N28\_B1&    60.35798 &   28.50189 &    2012-10-24 &               GAC051N24M1 &   51.07698 &   24.72406  \\ 
 2011-10-28  &          GAC\_105N29\_B1&   105.02926 &   29.77213 &    2012-10-24 &               GAC100N32B1 &  100.68762 &   32.55887  \\ 
 2011-11-08  &          GAC\_097N28\_B1&    97.59171 &   28.21211 &    2012-10-24 &               GAC100N32M1 &  100.68762 &   32.55887  \\ 
 2011-11-08  &          GAC\_122N29\_B1&   122.39659 &   29.09308 &    2012-10-25 &               GAC086N24B1 &   86.98389 &   24.68684  \\ 
 2011-11-09  &          GAC\_072N28\_B1&    72.31558 &   28.35083 &    2012-10-25 &               GAC086N24B2 &   86.98389 &   24.68684  \\ 
 2011-11-09  &          GAC\_101N28\_B1&   101.18941 &   28.97093 &    2012-10-25 &               GAC086N24M1 &   86.98389 &   24.68684  \\ 
 2011-11-09  &          GAC\_118N28\_B1&   118.33777 &   28.05199 &    2012-10-27 &               GAC090N26B1 &   90.89135 &   26.52913  \\ 
 2011-11-10  &          GAC\_089N28\_B1&    89.14070 &   28.94227 &    2012-10-27 &               GAC090N26B2 &   90.89135 &   26.52913  \\ 
 2011-11-10  &          GAC\_089N28\_B2&    89.14070 &   28.94227 &    2012-10-27 &               GAC102N27B1 &  102.29246 &   27.19037  \\ 
 2011-11-10  &          GAC\_089N28\_B3&    89.14070 &   28.94227 &    2012-10-27 &               GAC113N27B1 &  113.05368 &   27.12516  \\ 
 2011-11-10  &          GAC\_113N28\_B1&   113.63163 &   28.68658 &    2012-10-29 &               GAC078N26B1 &   78.08962 &   26.45461  \\ 
 2011-11-11  &          GAC\_082N29\_B1&    82.41940 &   29.18646 &    2012-10-29 &               GAC098N33B1 &   98.42774 &   33.02404  \\ 
 2011-11-12  &          GAC\_080N28\_B1&    80.84530 &   28.93676 &    2012-10-29 &               GAC113N26B1 &  113.98061 &   26.89574  \\ 
 2011-11-14  &          GAC\_087N27\_B1&    87.66559 &   27.50503 &    2012-10-29 &               GAC117N24B1 &  117.32855 &   24.48873  \\ 
 2011-11-14  &          GAC\_106N28\_B1&   106.85366 &   28.17669 &    2012-10-31 &               GAC097N26B1 &   97.23493 &   26.96746  \\ 
 2011-11-20  &   GAC\_063022N281243\_F1&    97.59171 &   28.21211 &    2012-10-31 &               GAC114N33B1 &  114.58637 &   33.18676  \\ 
 2011-11-20  &   GAC\_063022N281243\_M1&    97.59171 &   28.21211 &    2012-11-06 &          test\_055N28\_B1 &   55.32663 &   28.70276  \\ 
 2011-11-20  &   GAC\_080935N290534\_M1&   122.39659 &   29.09308 &    2012-11-06 &          test\_080N33\_B1 &   79.75012 &   33.74839  \\ 
 2011-11-23  &          GAC\_107N27\_B1&   108.98819 &   27.89742 &    2012-11-06 &          test\_080N33\_B2 &   79.75012 &   33.74839  \\ 
 2011-11-23  &          GAC\_107N27\_M1&   108.98819 &   27.89742 &    2012-11-06 &          test\_122N25\_B1 &  122.55461 &   25.84448  \\ 
 2011-11-24  &          GAC\_067N28\_M1&    72.31558 &   28.35083 &    2012-11-07 &               GAC049N32B1 &   49.14662 &   32.18402  \\ 
 2011-11-24  &          GAC\_106N28\_M1&   106.85366 &   28.17669 &    2012-11-07 &               GAC049N32M1 &   49.14662 &   32.18402  \\ 
 2011-11-26  &          GAC\_082N27\_M1&    87.66559 &   27.50503 &    2012-11-07 &               GAC069N25B1 &   69.84645 &   25.21827  \\ 
 2011-11-29  &          GAC\_096N28\_M1&   100.33707 &   28.19664 &    2012-11-07 &               GAC107N32B1 &  107.68018 &   32.61873  \\ 
 2011-11-30  &                    PB01B&    47.40309 &   29.07708 &    2012-11-07 &               GAC117N27B1 &  117.47942 &   27.36318  \\ 
 2011-12-03  &          GAC\_045N28\_B1&    45.87566 &   28.26991 &    2012-11-13 &               GAC073N32B1 &   73.87681 &   32.78394  \\ 
 2011-12-03  &          GAC\_045N28\_M1&    45.87566 &   28.26991 &    2012-11-13 &               GAC073N32M1 &   73.87681 &   32.78394  \\ 
 2011-12-03  &          GAC\_083N28\_M1&    89.14070 &   28.94227 &    2012-11-13 &               GAC102N32B1 &  102.42213 &   32.60676  \\ 
 2011-12-07  &          GAC\_106N28\_B1&   106.85366 &   28.17669 &    2012-11-13 &               GAC102N32M1 &  102.42213 &   32.60676  \\ 
 2011-12-11  &          GAC\_045N28\_B1&    45.87566 &   28.26991 &    2012-11-14 &               GAC098N33F1 &   98.42774 &   33.02404  \\ 
 2011-12-11  &          GAC\_065N28\_B1&    68.65830 &   28.96115 &    2012-11-14 &               GAC098N33M1 &   98.42774 &   33.02404  \\ 
 2011-12-11  &          GAC\_079N29\_B1&    82.52563 &   29.54832 &    2012-11-14 &               GAC121N33B1 &  121.03494 &   33.03074  \\ 
 2011-12-14  &          GAC\_078N28\_B1&    80.84530 &   28.93676 &    2012-11-17 &               GAC079N24B1 &   79.38239 &   24.00968  \\ 
 2011-12-14  &          GAC\_078N28\_M1&    80.84530 &   28.93676 &    2012-11-17 &               GAC079N24M1 &   79.38239 &   24.00968  \\ 
 2011-12-14  &          GAC\_105N29\_B1&   105.02926 &   29.77213 &    2012-11-19 &               GAC058N31F1 &   58.01852 &   31.16858  \\ 
 2011-12-15  &          GAC\_082N27\_B1&    87.66559 &   27.50503 &    2012-11-19 &               GAC121N33F1 &  121.03494 &   33.03074  \\ 
 2011-12-16  &                    PB03B&    56.15493 &   27.89745 &    2012-11-19 &               GAC121N33M1 &  121.03494 &   33.03074  \\ 
 2011-12-16  &                    PB03M&    56.15493 &   27.89745 &    2012-11-22 &               GAC049N27B1 &   49.98249 &   27.07113  \\ 
 2011-12-17  &   GAC\_034118N284209\_F1&    55.32663 &   28.70276 &    2012-11-22 &               GAC089N24F1 &   89.23383 &   24.24969  \\ 
 2011-12-18  &           GAC\_04h29\_B1&    61.75189 &   29.00130 &    2012-11-22 &               GAC089N24M1 &   89.23383 &   24.24969  \\ 
 2011-12-18  &           GAC\_04h29\_M1&    61.75189 &   29.00130 &    2012-11-23 &               GAC083N24B1 &   83.37816 &   24.62881  \\ 
 2011-12-19  &          GAC\_079N29\_M1&    82.41939 &   29.18646 &    2012-11-23 &               GAC104N26B1 &  104.69754 &   26.08108  \\ 
 2011-12-20  &          GAC\_067N28\_F1&    72.31558 &   28.35083 &    2012-11-23 &               GAC120N25B1 &  120.23281 &   25.39284  \\ 
 2011-12-21  &          GAC\_107N27\_B1&   108.98819 &   27.89742 &	   2012-11-25 &               GAC117N31B1 &  117.75082 &   31.61353  \\ 
 2011-12-22  &          GAC\_082N29\_M1&    82.52563 &   29.54832 &	   2012-12-04 &               GAC041N29B1 &   41.91085 &   29.67853  \\ 
 2011-12-23  &          GAC\_118N28\_F1&   118.33777 &   28.05199 &	   2012-12-04 &               GAC117N24M1 &  117.32855 &   24.48873  \\ 
 2011-12-23  &          GAC\_118N28\_M1&   118.33777 &   28.05199 &	   2012-12-05 &               GAC065N31M1 &   65.04145 &   31.95317  \\ 
 2011-12-24  &          GAC\_068N28\_F1&    68.65830 &   28.96115 &	   2012-12-06 &          test\_091N23\_B1 &   91.70276 &   23.63860  \\ 
 2011-12-24  &          GAC\_068N28\_M1&    68.65830 &   28.96115 &	   2012-12-06 &          test\_091N23\_M1 &   91.70276 &   23.63860  \\ 
 2011-12-24  &          GAC\_106N28\_B1&   106.85366 &   28.17669 &	   2012-12-06 &          test\_111N36\_B1 &  111.07162 &   36.31091  \\ 
 2011-12-25  &          GAC\_089N28\_B3&    89.14070 &   28.94227 &	   2012-12-06 &          test\_128N24\_B1 &  127.87716 &   24.08111  \\ 
 2011-12-25  &          GAC\_089N28\_M3&    89.14070 &   28.94227 &	   2012-12-06 &          test\_128N24\_B2 &  127.87716 &   24.08111  \\ 
 2011-12-25  &          GAC\_113N28\_M1&   113.63163 &   28.68658 &	   2012-12-08 &               GAC074N27B1 &   74.97396 &   27.32560  \\ 
\end{tabular} 
\end{small}
\end{table*}  

\setcounter{table}{0}
\begin{table*}
\caption{Continued. }  
\setlength{\tabcolsep}{0.9ex}
\centering
\begin{small}
\begin{tabular}{cccccccc}
\hline
\hline
Date & Plate & RA & DEC & Date & Plate & RA & DEC  \\
    &  & [deg] & [deg]    & & & [deg] & [deg]  \\
\hline
 2011-12-26  &          GAC\_047N29\_B1&    47.40309 &   29.07708 &	   2012-12-08 &               GAC074N27M1 &   74.97396 &   27.32560  \\ 
 2011-12-26  &          GAC\_091N29\_B1&    91.59353 &   29.51241 &	   2012-12-09 &               GAC084N26B1 &   84.73905 &   26.61807  \\ 
 2011-12-26  &          GAC\_091N29\_M1&    91.59353 &   29.51241 &	   2012-12-09 &               GAC084N26M1 &   84.73905 &   26.61807  \\ 
 2011-12-27  &          GAC\_105N29\_M1&   105.02926 &   29.77213 &	   2012-12-09 &               GAC114N32B1 &  114.97536 &   32.00973  \\ 
 2011-12-28  &          GAC\_108N27\_M1&   108.98819 &   27.89742 &	   2012-12-21 &               GAC120N25M1 &  120.23281 &   25.39284  \\ 
 2011-12-31  &                    PA06F&    68.65830 &   28.96115 &	   2012-12-22 &               GAC108N24B1 &  108.10991 &   24.12859  \\ 
 2011-12-31  &                    PA06M&    68.65830 &   28.96115 &	   2013-01-04 &          test\_076N27\_B1 &   76.15819 &   27.69602  \\ 
 2012-01-02  &                    PB06B&    82.41939 &   29.18646 &	   2013-01-04 &          test\_076N27\_M1 &   76.15819 &   27.69602  \\ 
 2012-01-02  &                    PB06M&    82.41939 &   29.18646 &	   2013-01-04 &          test\_094N28\_B1 &   94.99571 &   28.42676  \\ 
 2012-01-03  &          GAC\_068N28\_B1&    68.65830 &   28.96115 &	   2013-01-04 &          test\_126N31\_B1 &  126.02454 &   31.30104  \\ 
 2012-01-03  &          GAC\_101N28\_M1&   100.33707 &   28.19664 &	   2013-01-05 &               GAC087N32B1 &   87.85729 &   32.12469  \\ 
 2012-01-04  &          GAC\_063N29\_B1&    63.79331 &   29.90206 &	   2013-01-05 &               GAC087N32M1 &   87.85729 &   32.12469  \\ 
 2012-01-04  &          GAC\_106N28\_M1&   106.85366 &   28.17669 &	   2013-01-06 &               GAC096N32B1 &   96.30344 &   32.27176  \\ 
 2012-01-11  &          GAC\_050N29\_B1&    50.08484 &   29.04846 &	   2013-01-06 &               GAC096N32M1 &   96.30344 &   32.27176  \\ 
 2012-01-12  &          GAC\_045N28\_B1&    45.87566 &   28.26991 &	   2013-01-07 &               GAC085N33M1 &   85.74273 &   33.31455  \\ 
 2012-01-12  &          GAC\_100N28\_B1&   100.33707 &   28.19664 &	   2013-01-07 &               GAC122N25B1 &  122.55461 &   25.84448  \\ 
 2012-01-13  &          GAC\_061N29\_B1&    61.75189 &   29.00130 &	   2013-01-07 &               GAC122N25M1 &  122.55461 &   25.84448  \\ 
 2012-01-14  &                   PA09B1&    89.14070 &   28.94227 &	   2013-01-08 &               GAC086N24M2 &   86.98389 &   24.68684  \\ 
 2012-01-14  &                   PA09M1&    89.14070 &   28.94227 &	   2013-01-09 &               GAC085N31B1 &   85.14961 &   31.35820  \\ 
 2012-01-14  &                    PB02B&    52.19438 &   30.37534 &	   2013-01-09 &               GAC085N31M1 &   85.14961 &   31.35820  \\ 
 2012-01-15  &                    PA01B&    45.87566 &   28.26991 &	   2013-01-11 &               GAC094N27M1 &   94.58662 &   27.21015  \\ 
 2012-01-15  &                   PA09B1&    89.14070 &   28.94227 &	   2013-01-12 &               GAC054N25M1 &   54.29599 &   25.99110  \\ 
 2012-01-15  &                    PB05B&    70.09451 &   29.97223 &	   2013-01-12 &               GAC078N26M1 &   78.08962 &   26.45461  \\ 
 2012-01-21  &          GAC\_060N28\_F1&    60.35798 &   28.50189 &	   2013-01-12 &               GAC109N30M1 &  109.51705 &   30.95587  \\ 
 2012-01-21  &          GAC\_100N28\_M1&   100.33707 &   28.19664 &	   2013-01-13 &               GAC080N33B2 &   79.75012 &   33.74839  \\ 
 2012-01-22  &          GAC\_063N29\_M1&    63.79331 &   29.90206 &	   2013-01-13 &               GAC080N33M1 &   79.75012 &   33.74839  \\ 
 2012-01-22  &          GAC\_089N28\_F1&    89.14070 &   28.94227 &	   2013-01-16 &               GAC045N26B1 &   45.47559 &   26.46235  \\ 
 2012-01-23  &          GAC\_061N29\_M1&    61.75189 &   29.00130 &	   2013-01-16 &               GAC055N32B1 &   55.61516 &   32.93923  \\ 
 2012-01-23  &          GAC\_089N28\_F2&    89.14070 &   28.94227 &	   2013-01-16 &               GAC065N31B1 &   65.04145 &   31.95317  \\ 
 2012-01-24  &          GAC\_060N28\_M1&    60.35798 &   28.50189 &	   2013-01-17 &               GAC040N27B1 &   40.86297 &   27.70715  \\ 
 2012-01-24  &          GAC\_080N28\_M1&    80.84530 &   28.93676 &	   2013-01-17 &               GAC105N24B1 &  105.60325 &   24.21545  \\ 
 2012-01-24  &          GAC\_108N27\_M1&   108.98819 &   27.89742 &	   2013-01-29 &               GAC058N25B1 &   57.85539 &   25.16293  \\ 
 2012-01-25  &          GAC\_052N30\_M1&    52.19438 &   30.37534 &	   2013-02-01 &               GAC076N33B1 &   76.53483 &   33.91869  \\ 
 2012-01-25  &          GAC\_097N28\_F1&    97.59171 &   28.21211 &	   2013-02-04 &               GAC046N25B1 &   46.36121 &   25.25517  \\ 
 2012-01-26  &          GAC\_055N28\_M1&    55.32663 &   28.70276 &	   2013-02-07 &               GAC054N25B1 &   54.29599 &   25.99110  \\ 
 2012-01-26  &          GAC\_089N28\_M2&    89.14070 &   28.94227 &	   2013-02-08 &               GAC046N25B1 &   46.36121 &   25.25517  \\ 
 2012-01-26  &          GAC\_122N29\_M1&   122.39659 &   29.09308 &	   2013-02-08 &               GAC062N26B1 &   62.70775 &   26.48095  \\ 
 2012-01-29  &                    PA01M&    45.87566 &   28.26991 &	   2013-02-08 &               GAC109N31B1 &  109.37725 &   31.69805  \\ 
 2012-01-29  &                    PA11F&   106.85366 &   28.17669 &	   2013-02-08 &               GAC109N31M1 &  109.37725 &   31.69805  \\ 
 2012-01-29  &                    PB05M&    70.09451 &   29.97223 &	   2013-02-09 &               GAC072N28B1 &   72.31558 &   28.35083  \\ 
 2012-01-31  &                    PB08M&    91.59353 &   29.51241 &	   2013-02-09 &               GAC094N27B1 &   94.58662 &   27.21015  \\ 
 2012-01-31  &                    PB09B&   100.33707 &   28.19664 &	   2013-02-10 &               GAC110N25B1 &  110.86880 &   25.05053  \\ 
 2012-02-14  &                    PA08F&   124.98256 &   39.88470 &	   2013-02-14 &               GAC053N32B1 &   53.75501 &   32.01677  \\ 
 2012-02-16  &          GAC\_106N28\_F1&   106.85366 &   28.17669 &	   2013-02-14 &               GAC114N32B1 &  114.97536 &   32.00973  \\ 
 2012-03-13  &                    PA13B&   129.34213 &   28.29440 &	   2013-02-15 &               GAC076N30B1 &   76.06069 &   30.49459  \\ 
 2012-03-14  &                    PB09B&   100.33707 &   28.19664 &	   2013-02-18 &               GAC053N32B1 &   53.75501 &   32.01677  \\ 
 2012-10-03  &            GAC080N33B101&    79.75012 &   33.74839 &	   2013-03-04 &           GAC\_098N33\_B1 &   98.42774 &   33.02404  \\ 
 2012-10-05  &              GAC054N25B1&    54.29599 &   25.99110 &	   2013-03-04 &           GAC\_128N36\_B1 &  128.34054 &   36.43643  \\ 
 2012-10-05  &              GAC067N27B1&    67.33264 &   27.40422 &	   2013-03-04 &           GAC\_128N36\_B2 &  128.34054 &   36.43643  \\ 
 2012-10-05  &              GAC089N24B1&    89.23383 &   24.24969 &	   2013-03-05 &               GAC078N26B1 &   78.08962 &   26.45461  \\ 
 2012-10-06  &              GAC085N33B1&    85.74273 &   33.31455 &	   2013-03-05 &               GAC108N24B1 &  108.10991 &   24.12859  \\ 
 2012-10-06  &              GAC085N33B2&    85.74273 &   33.31455 &	   2013-03-06 &               GAC089N28B1 &   89.14070 &   28.94227  \\ 
 2012-10-07  &              GAC081N30B1&    81.78447 &   30.20860 &	   2013-03-06 &               GAC089N28B2 &   89.14070 &   28.94227  \\ 
 2012-10-07  &              GAC081N30B2&    81.78447 &   30.20860 &	   2013-03-07 &               GAC082N33B1 &   82.00390 &   33.76370  \\ 
 2012-10-12  &              GAC056N24B1&    56.20090 &   24.28947 &	   2013-04-03 &          test\_114N22\_B1 &  114.41732 &   22.34006  \\ 
 2012-10-12  &              GAC056N24F1&    56.20090 &   24.28947 &                   &                           &            &             \\
\hline        
\end{tabular} 
\end{small}
\end{table*}

\begin{table*}
\caption{\label{t-plates} Names, coordinates, plate-spectrograph-fibre
  identifiers, XSTPS-GAC magnitudes  and stellar parameters (effective
  temperature, surface  gravity and mass)  of the 92 DA  LSS-GAC white
  dwarfs  identified  in  this work.}  
\setlength{\tabcolsep}{0.5ex}
\centering
\begin{small}
\begin{tabular}{ccccccccccccccc}
\hline
\hline
Jname              &    RA      &     DEC             & plateid  &  spid & fiberid & $g$ & $r$ & $i$ &  \Teff  &  err & $\log(g)$ & err & $M$ & err\\ 
                   &   [deg]    &     [deg]           &          &       &         &   &   &   &   [k]   &      &  [dex]    &     & [M$_\odot$]  &    \\
\hline
J062647.53+264552.1 &  96.69805   &   26.76446 &        GAC\_097N28\_F1 &  5 & 220 & 17.53 & 17.64 & 18.03 &   27425 &    562  &  8.15 &   0.10 &   0.73 &   0.06 \\
J070057.53+284310.1 & 105.23969   &   28.71946 &        GAC\_105N29\_M1 &  5 & 129 & 17.37 & 16.98 & 17.23 &   13173 &    462  &  8.27 &   0.16 &   0.78 &   0.11 \\
J074722.19+295015.3 & 116.84244   &   29.83757 &        GAC\_118N28\_F1 & 16 &  27 & 18.19 & 17.83 & 18.05 &   15601 &   1150  &  7.78 &   0.28 &   0.50 &   0.15 \\
J074850.71+301003.5 & 117.21131   &   30.16765 &        GAC\_118N28\_F1 & 16 & 228 & 18.27 & 17.89 & 18.11 &     - &     - &     - &     - &     - &     - \\
J075106.48+301727.0 & 117.77698   &   30.29082 &        GAC\_118N28\_M1 & 11 & 169 & 15.65 & 15.92 & 16.39 &   32222 &    401  &  8.04 &   0.10 &   0.67 &   0.06 \\
J074742.05+280945.6 & 116.92519   &   28.16266 &        GAC\_118N28\_F1 &  3 &  60 & 17.83 & 17.43 & 17.69 &   13234 &    286  &  8.18 &   0.09 &   0.72 &   0.06 \\
J075251.35+271513.9 & 118.21396   &   27.25385 &        GAC\_118N28\_M1 &  5 & 171 & 16.72 & 16.73 & 17.15 &   23343 &    737  &  7.78 &   0.11 &   0.52 &   0.06 \\
J034037.94+230304.2 &  55.15809   &   23.05117 &            GAC056N24F1 &  5 & 224 & 18.07 & 18.35 & 18.35 &     - &     - &     - &     - &     - &     - \\
J051002.11+231541.0 &  77.50881   &   23.26140 &            GAC079N24B1 & 10 &  62 & 14.70 & 15.26 & 15.46 &   11134 &    131  &  7.89 &   0.11 &   0.56 &   0.07 \\
J060027.42+234311.2 &  90.11425   &   23.71979 &            GAC089N24M1 &  8 &   7 & 16.98 & 16.98 & 16.98 &    8205 &    113  &  7.99 &   0.22 &   0.59 &   0.15 \\
J033633.53+234938.7 &  54.13971   &   23.82742 &            GAC056N24F1 & 10 &  57 & 17.63 & 17.93 & 17.96 &   30419 &   1165  &  7.88 &   0.30 &   0.58 &   0.16 \\
J073145.10+235352.9 & 112.93792   &   23.89802 &            GAC110N25B1 &  7 & 144 & 16.00 & 16.22 & 16.45 &   11736 &    372  &  7.75 &   0.27 &   0.47 &   0.16 \\
J033900.07+242510.5 &  54.75028   &   24.41957 &            GAC056N24F1 &  3 &  53 & 18.01 & 18.49 & 18.52 &   17304 &    953  &  7.88 &   0.22 &   0.55 &   0.12 \\
J055727.92+243558.8 &  89.36633   &   24.59967 &            GAC086N24M1 &  6 &  56 & 16.47 & 16.97 & 17.27 &   29387 &    567  &  8.04 &   0.13 &   0.67 &   0.08 \\
J080230.40+244922.6 & 120.62665   &   24.82295 &            GAC120N25M1 &  4 & 119 & 17.66 & 17.75 & 17.87 &    9070 &    222  &  8.15 &   0.34 &   0.69 &   0.21 \\
J054335.91+250410.8 &  85.89961   &   25.06966 &            GAC084N26M1 &  7 & 122 & 15.90 & 16.44 & 16.70 &   68101 &   4213  &  7.37 &   0.20 &   0.51 &   0.06 \\
J034422.26+251453.3 &  56.09274   &   25.24813 &            GAC058N25B1 & 14 &  66 & 15.39 & 15.71 & 15.85 &    7931 &    101  &  7.74 &   0.29 &   0.45 &   0.18 \\
J062159.50+252335.9 &  95.49790   &   25.39331 &            GAC094N27M1 &  7 &  21 & 17.56 & 17.62 & 17.70 &   11728 &    651  &  8.25 &   0.31 &   0.76 &   0.19 \\
J032817.13+252853.5 &  52.07136   &   25.48152 &            GAC051N24M1 &  9 &  36 & 16.57 & 16.98 & 17.15 &   14132 &    944  &  8.00 &   0.16 &   0.61 &   0.10 \\
J032854.06+252626.3 &  52.22527   &   25.44064 &            GAC051N24M1 &  9 &  35 & 18.06 & 17.73 & 17.54 &     - &     - &     - &     - &     - &     - \\
J054613.53+255031.7 &  86.55636   &   25.84214 &        GAC\_082N27\_M1 &  2 & 249 & 17.33 & 17.62 & 17.78 &     - &     - &     - &     - &     - &     - \\
J070950.16+255303.6 & 107.45899   &   25.88434 &        GAC\_106N28\_M1 &  1 & 184 & 17.10 & 17.63 & 17.94 &   77295 &  10207  &  7.64 &   0.42 &   0.61 &   0.16 \\
J053727.46+260611.3 &  84.36441   &   26.10313 &            GAC084N26M1 &  4 & 135 & 16.76 & 17.17 & 17.33 &   15782 &    488  &  7.95 &   0.11 &   0.59 &   0.06 \\
J064452.84+260947.7 & 101.22017   &   26.16326 &        GAC\_100N28\_B1 &  7 & 168 & 15.48 & 15.98 & 16.22 &   16910 &    671  &  7.60 &   0.16 &   0.42 &   0.07 \\
J071223.81+260933.4 & 108.09919   &   26.15928 &        GAC\_106N28\_M1 &  7 & 226 & 16.85 & 17.24 & 17.41 &   14132 &    645  &  8.00 &   0.15 &   0.61 &   0.09 \\
J055046.50+261220.3 &  87.69377   &   26.20563 &        GAC\_082N27\_B1 &  5 &  18 & 15.13 & 15.64 & 15.91 &   21289 &    814  &  7.60 &   0.14 &   0.43 &   0.06 \\
J063532.49+261958.6 &  98.88537   &   26.33295 & GAC\_063022N281243\_M1 &  7 & 221 & 16.63 & 17.17 & 17.47 &   34526 &    928  &  8.01 &   0.18 &   0.66 &   0.11 \\
J063828.24+263359.6 &  99.61767   &   26.56656 &        GAC\_100N28\_B1 &  2 &  52 & 16.03 & 15.60 & 15.48 &     - &     - &     - &     - &     - &     - \\
J063856.01+263022.5 &  99.73339   &   26.50626 &        GAC\_100N28\_B1 &  2 &  51 & 14.85 & 14.35 & 14.22 &     - &     - &     - &     - &     - &     - \\
J025737.25+264047.9 &  44.40520   &   26.67997 &        GAC\_045N28\_M1 &  2 & 143 & 16.91 & 17.00 & 17.16 &     - &     - &     - &     - &     - &     - \\
J063810.91+264040.9 &  99.54544   &   26.67802 &        GAC\_100N28\_B1 &  2 &  54 & 16.32 & 15.59 & 15.28 &     - &     - &     - &     - &     - &     - \\
J063919.88+264102.6 &  99.83283   &   26.68405 &        GAC\_100N28\_B1 &  2 &  56 & 15.97 & 15.58 & 15.40 &     - &     - &     - &     - &     - &     - \\
J070716.11+263857.5 & 106.81713   &   26.64930 &        GAC\_106N28\_M1 &  1 & 154 & 17.63 & 18.07 & 18.20 &   14899 &    783  &  7.29 &   0.19 &   0.31 &   0.06 \\
J034605.50+264348.3 &  56.52290   &   26.73008 & GAC\_034118N284209\_F1 &  7 & 240 & 17.54 & 17.95 & 18.26 &   22550 &   2127  &  7.95 &   0.35 &   0.60 &   0.21 \\
J070755.01+265103.0 & 106.97921   &   26.85082 &        GAC\_106N28\_B1 &  5 &  12 & 15.53 & 15.86 & 16.01 &   15071 &    952  &  7.89 &   0.20 &   0.55 &   0.12 \\
J070706.33+265756.7 & 106.77636   &   26.96574 &        GAC\_106N28\_M1 &  5 & 138 & 17.19 & 17.63 & 17.83 &   16910 &    564  &  8.21 &   0.14 &   0.74 &   0.09 \\
J064743.70+270906.2 & 101.93207   &   27.15172 &        GAC\_096N28\_M1 &  6 &  25 & 16.00 & 16.59 & 16.90 &   41991 &   1207  &  8.21 &   0.12 &   0.79 &   0.07 \\
J034409.58+271507.3 &  56.03992   &   27.25202 &                  PB03M &  5 & 172 & 16.02 & 16.52 & 16.75 &   19868 &    246  &  7.78 &   0.05 &   0.51 &   0.02 \\
J054447.48+272032.0 &  86.19783   &   27.34223 &        GAC\_082N27\_M1 &  3 &  76 & 17.08 & 16.93 & 16.90 &    6459 &     56  &  7.75 &   0.14 &   0.45 &   0.07 \\
J071520.95+273433.4 & 108.83731   &   27.57594 &        GAC\_106N28\_M1 &  6 & 192 & 17.93 & 18.01 & 18.12 &     - &     - &     - &     - &     - &     - \\
J061540.92+275202.0 &  93.92048   &   27.86723 &            GAC094N27M1 & 15 & 229 & 16.47 & 16.91 & 17.20 &   46042 &   1008  &  8.82 &   0.11 &   1.13 &   0.05 \\
J035123.16+280256.6 &  57.84651   &   28.04905 & GAC\_034118N284209\_F1 &  6 & 169 & 17.75 & 18.11 & 18.51 &   32222 &    862  &  8.90 &   0.19 &   1.16 &   0.09 \\
J061000.28+281426.9 &  92.50117   &   28.24081 &       test\_094N28\_B1 & 10 & 131 & 14.33 & 14.74 & 15.10 &   18542 &     62  &  7.97 &   0.02 &   0.60 &   0.01 \\
J042945.25+282224.2 &  67.43856   &   28.37338 &        GAC\_068N28\_B1 &  3 & 248 & 14.60 & 15.21 & 15.52 &   39642 &    699  &  8.06 &   0.09 &   0.70 &   0.05 \\
J074741.42+282104.9 & 116.92258   &   28.35137 &        GAC\_118N28\_F1 &  3 &  54 & 18.08 & 18.40 & 18.64 &   13284 &   2679  &  8.10 &   0.43 &   0.67 &   0.27 \\
J080748.88+282626.4 & 121.95366   &   28.44067 & GAC\_080935N290534\_M1 &  5 &  55 & 17.79 & 17.93 & 18.02 &    9356 &    256  &  7.97 &   0.51 &   0.58 &   0.32 \\
J052147.24+283532.5 &  80.44682   &   28.59236 &        GAC\_080N28\_M1 &  4 & 198 & 17.50 & 17.73 & 17.86 &   10780 &    199  &  8.29 &   0.15 &   0.79 &   0.09 \\
J030406.41+285143.2 &  46.02670   &   28.86200 &        GAC\_045N28\_M1 &  4 & 206 & 16.82 & 17.19 & 17.39 &     - &     - &     - &     - &     - &     - \\
J030214.72+285707.4 &  45.56134   &   28.95206 &        GAC\_045N28\_M1 & 15 & 183 & 17.21 & 17.60 & 17.80 &   18120 &   1667  &  7.51 &   0.35 &   0.39 &   0.15 \\
J053931.86+285456.7 &  84.88277   &   28.91574 &        GAC\_079N29\_M1 &  6 & 167 & 17.39 & 16.64 & 16.17 &     - &     - &     - &     - &     - &     - \\
J042334.22+290205.7 &  65.89260   &   29.03492 &        GAC\_068N28\_F1 & 14 &  44 & 18.10 & 18.32 & 18.38 &   24443 &   2180  &  8.06 &   0.35 &   0.67 &   0.21 \\
J080800.00+290152.6 & 122.00001   &   29.03127 & GAC\_080935N290534\_M1 &  4 & 155 & 16.64 & 17.19 & 17.48 &   24163 &    600  &  7.78 &   0.09 &   0.52 &   0.04 \\
J033134.40+291321.4 &  52.89335   &   29.22260 &        GAC\_055N28\_M1 & 14 & 111 & 17.14 & 17.40 & 17.55 &   14393 &   1093  &  7.56 &   0.26 &   0.39 &   0.12 \\
J040403.73+291703.8 &  61.01555   &   29.28440 &        GAC\_060N28\_M1 &  9 & 165 & 17.73 & 17.80 & 17.73 &     - &     - &     - &     - &     - &     - \\
J040842.17+292130.3 &  62.17569   &   29.35843 &        GAC\_063N29\_M1 & 10 & 223 & 17.06 & 17.01 & 16.97 &    7659 &    113  &  7.95 &   0.23 &   0.56 &   0.15 \\
J071004.83+292402.8 & 107.52014   &   29.40079 &        GAC\_107N27\_B1 & 16 & 178 & 15.29 & 15.62 & 15.87 &   14089 &    732  &  7.94 &   0.18 &   0.57 &   0.11 \\
J042435.27+293651.7 &  66.14697   &   29.61437 &        GAC\_068N28\_M1 & 14 & 104 & 17.29 & 17.52 & 17.69 &   36572 &   1719  &  8.22 &   0.30 &   0.79 &   0.18 \\
J054658.08+293633.4 &  86.74200   &   29.60928 &                 PA09M1 & 14 & 111 & 17.23 & 17.29 & 17.16 &   28064 &    789  &  7.84 &   0.16 &   0.56 &   0.08 \\
J065740.34+300915.8 & 104.41810   &   30.15438 &        GAC\_105N29\_M1 &  3 & 134 & 17.14 & 17.08 & 17.06 &    7839 &     43  &  7.90 &   0.11 &   0.54 &   0.07 \\
J071037.96+301146.4 & 107.65815   &   30.19622 &            GAC109N30M1 & 10 &  26 & 17.23 & 17.30 & 17.38 &     - &     - &     - &     - &     - &     - \\
J030128.01+301536.6 &  45.36671   &   30.26016 &        GAC\_045N28\_B1 & 11 &  19 & 14.75 & 15.18 & 15.36 &     - &     - &     - &     - &     - &     - \\
J071011.40+303041.7 & 107.54751   &   30.51158 &        GAC\_106N28\_M1 & 11 &  27 & 16.46 & 16.84 & 17.10 &   18756 &    594  &  8.33 &   0.12 &   0.82 &   0.08 \\
\end{tabular} 
\end{small}
\end{table*}
  
\setcounter{table}{1}
\begin{table*}
\caption{Continued.}  
\setlength{\tabcolsep}{0.5ex}
\centering
\begin{small}
\begin{tabular}{ccccccccccccccc}
\hline
\hline
Jname              &    RA      &     DEC             & plateid  &  spid & fiberid & $g$ & $r$ & $i$ &  \Teff  &  err & $\log(g)$ & err & $M$ & err\\ 
                   &   [deg]    &     [deg]           &          &       &         &   &   &   &   [k]   &      &  [dex]    &     & [M$_\odot$]  &    \\
\hline
J051624.54+303910.3 &  79.10227   &   30.65286 &            GAC080N32F1 &  2 & 118 & 17.84 & 18.14 & 18.33 &   21535 &   2125  &  8.46 &   0.37 &   0.91 &   0.21 \\
J060113.75+304014.3 &  90.30730   &   30.67064 &                 PA09M1 & 12 &  66 & 17.49 & 17.66 & 17.73 &   10939 &    330  &  8.12 &   0.31 &   0.68 &   0.20 \\
J052038.36+304822.6 &  80.15984   &   30.80629 &        GAC\_078N28\_B1 & 16 & 139 & 15.38 & 15.68 & 15.88 &   15601 &    395  &  7.87 &   0.09 &   0.54 &   0.05 \\
J033149.68+305944.9 &  52.95702   &   30.99581 &        GAC\_052N30\_M1 &  9 & 224 & 16.89 & 17.41 & 17.67 &   19868 &    306  &  8.46 &   0.06 &   0.91 &   0.04 \\
J080039.67+305537.3 & 120.16529   &   30.92703 &            GAC121N33M1 &  2 & 118 & 17.44 & 17.84 & 18.18 &   25302 &   1399  &  7.58 &   0.20 &   0.44 &   0.08 \\
J031430.60+310301.4 &  48.62749   &   31.05040 &            GAC049N32M1 &  5 &  42 & 17.20 & 17.61 & 17.82 &   18120 &   1669  &  8.03 &   0.38 &   0.64 &   0.23 \\
J072438.30+310729.0 & 111.15957   &   31.12471 &            GAC109N30M1 &  9 & 238 & 17.35 & 17.51 & 17.69 &   10772 &    381  &  8.79 &   0.31 &   1.09 &   0.11 \\
J075724.50+311310.3 & 119.35207   &   31.21953 &            GAC121N33F1 &  2 & 147 & 17.73 & 18.19 & 18.41 &   20098 &    741  &  7.85 &   0.14 &   0.54 &   0.07 \\
J063058.95+312344.1 &  97.74564   &   31.39559 &            GAC096N32M1 &  6 & 239 & 17.24 & 17.53 & 17.76 &   15071 &    803  &  7.72 &   0.19 &   0.47 &   0.10 \\
J055234.17+312401.6 &  88.14237   &   31.40045 &            GAC090N33B1 &  2 &  89 & 14.64 & 15.00 & 15.22 &   14228 &    176  &  8.15 &   0.03 &   0.70 &   0.02 \\
J082705.52+313008.2 & 126.77302   &   31.50228 &       test\_126N31\_B1 &  9 &  22 & 15.36 & 15.92 & 16.32 &   71310 &   4547  &  7.44 &   0.22 &   0.54 &   0.06 \\
J071603.19+315711.0 & 109.01331   &   31.95305 &            GAC109N30M1 & 15 &  11 & 17.72 & 17.91 & 17.86 &     - &     - &     - &     - &     - &     - \\
J080602.45+315431.5 & 121.51019   &   31.90875 &            GAC121N33M1 &  8 & 185 & 17.25 & 17.80 & 18.11 &   22811 &   1761  &  7.78 &   0.28 &   0.52 &   0.15 \\
J062856.67+320303.9 &  97.23611   &   32.05109 &            GAC098N33F1 &  5 & 193 & 17.77 & 18.31 & 18.75 &   36572 &   1334  &  7.71 &   0.24 &   0.53 &   0.11 \\
J071451.13+320407.9 & 108.71306   &   32.06885 &            GAC109N30M1 & 15 &  76 & 15.89 & 16.44 & 16.79 &   49906 &   2195  &  8.49 &   0.18 &   0.97 &   0.09 \\
J053712.39+321502.9 &  84.30162   &   32.25081 &            GAC085N31M1 & 15 & 243 & 16.70 & 16.96 & 17.19 &   16525 &    275  &  7.66 &   0.06 &   0.44 &   0.03 \\
J042006.39+323305.0 &  65.02661   &   32.55138 &            GAC065N31B1 &  4 & 216 & 15.76 & 16.20 & 16.43 &   17912 &    544  &  8.06 &   0.11 &   0.65 &   0.07 \\
J063100.30+324453.8 &  97.75127   &   32.74827 &            GAC100N32M1 & 14 &  38 & 16.16 & 16.76 & 17.09 &   64291 &   7505  &  7.70 &   0.41 &   0.60 &   0.18 \\
J031336.68+325107.9 &  48.40285   &   32.85219 &            GAC049N32M1 & 15 & 248 & 17.29 & 17.40 & 17.52 &   10642 &    294  &  8.43 &   0.26 &   0.87 &   0.14 \\
J031448.30+324916.3 &  48.70124   &   32.82120 &            GAC049N32M1 & 15 & 195 & 16.70 & 17.12 & 17.32 &   25595 &    915  &  8.24 &   0.13 &   0.78 &   0.08 \\
J054709.65+324843.7 &  86.79021   &   32.81214 &            GAC085N33M1 &  8 &   2 & 16.72 & 16.31 & 16.13 &     - &     - &     - &     - &     - &     - \\
J061736.85+325732.0 &  94.40353   &   32.95888 &            GAC096N32M1 & 14 & 173 & 16.49 & 16.97 & 17.20 &   35331 &    576  &  7.88 &   0.12 &   0.60 &   0.06 \\
J080006.18+325738.1 & 120.02574   &   32.96058 &            GAC121N33F1 &  3 & 179 & 18.15 & 18.45 & 18.63 &   14393 &   1410  &  8.44 &   0.20 &   0.89 &   0.13 \\
J030945.71+330025.0 &  47.44047   &   33.00694 &            GAC049N32M1 & 14 &  79 & 16.78 & 17.09 & 17.30 &   14899 &   1087  &  8.62 &   0.18 &   1.00 &   0.10 \\
J075321.55+334308.1 & 118.33978   &   33.71892 &            GAC121N33F1 & 14 & 103 & 17.98 & 18.41 & 18.68 &   36996 &   2504 &   7.07 &   0.41 &   0.34 &   0.11 \\
J055407.71+340348.0 &  88.53214   &   34.06334 &            GAC087N32M1 & 11 &  71 & 17.59 & 17.62 & 17.76 &   10409 &    188  &  7.19 &   0.40 &   0.27 &   0.17 \\
J055019.78+350006.5 &  87.58242   &   35.00180 &            GAC085N33M1 & 12 & 184 & 16.31 & 16.64 & 16.91 &   18542 &    404  &  7.70 &   0.08 &   0.47 &   0.04 \\
J051940.72+355137.1 &  79.91965   &   35.86031 &            GAC080N33M1 & 11 & 178 & 17.27 & 17.45 & 17.68 &   10656 &    174  &  8.08 &   0.21 &   0.65 &   0.14 \\
J083400.13+365921.7 & 128.50056   &   36.98935 &        GAC\_128N36\_B2 &  4 & 206 & 16.05 & 16.09 & 16.14 &    9065 &     67  &  7.93 &   0.12 &   0.56 &   0.08 \\
J084218.71+374859.4 & 130.57797   &   37.81649 &        GAC\_128N36\_B2 & 12 & 116 & 16.05 & 16.46 & 16.76 &   22037 &   1183  &  8.42 &   0.20 &   0.88 &   0.12 \\
\hline        
\end{tabular} 
\end{small}
\end{table*}  

\end{document}